\documentclass{aa}   
\usepackage{graphicx}
\usepackage{txfonts}
\begin{document}
\title{Neutral material around the B[e] supergiant star LHA\,115-S\,65\,\thanks{Based on observations done with the 1.52m and 2.2m telescope at the
European Southern Observatory (La Silla, Chile) under the agreement with the
Observat\'orio Nacional-MCT (Brazil) and under programs 076.D-0609(A) and 080.A-9200(A)}}

\subtitle{An outflowing disk or a detached Keplerian rotating disk?}

\author{M. Kraus\inst{1}, M. Borges Fernandes\inst{2,3}, \and F.X. de Ara\'ujo\inst{3}\thanks{We have to report with great sadness that our friend and colleague, Francisco Xavier de Ara\'ujo, deceased before this work could be finished}
  }
                                                                                
\institute{Astronomick\'y \'ustav, Akademie v\v{e}d \v{C}esk\'e republiky, Fri\v{c}ova 298, 251\,65 Ond\v{r}ejov, Czech Republic\\
\email{kraus@sunstel.asu.cas.cz}
 \and
UMR 6525 H. Fizeau, Univ. Nice Sophia Antipolis, CNRS, Observatoire de
la C\^{o}te d'Azur, Av. Copernic, F-06130 Grasse, France\\
\email{Marcelo.Borges@obs-azur.fr}
 \and
   Observat\'orio Nacional, Rua General Jos\'e Cristino 77, 20921-400 S\~ao Cristov\~ao, Rio de Janeiro, Brazil\\
\email{borges@on.br}
      }

\date{Received ; accepted }

\abstract
%Context
{
B[e] supergiants are surrounded by large amounts of hydrogen neutral material, 
traced by the emission in the optical [O{\sc i}] lines. This neutral 
material is most plausibly located within their dense, cool circumstellar 
disks, which are
formed from the (probably non-spherically symmetric) wind material released by 
the star. Neither the formation mechanism nor the resulting structure 
and internal kinematics of these disks (or disk-like outflows) are well 
known. However, rapid rotation, lifting the material from the equatorial 
surface region, seems to play a fundamental role.
}
%Aims
{
The B[e] supergiant LHA\,115-S\,65 (in short: S\,65) in the Small 
Magellanic Cloud is one of the two most rapidly rotating B[e] stars known.
Its almost edge-on orientation allows a detailed kinematical study of its
optically thin forbidden emission lines. With a focus on the rather strong
[O{\sc i}] lines, we intend to test the two plausible disk scenarios: the 
outflowing and the Keplerian rotating disk.
}
%Methods
{
Based on high- and low-resolution optical spectra, we investigate the 
density and temperature structure in those disk regions that are traced by the 
[O{\sc i}] emission to constrain the disk sizes and mass fluxes 
needed to explain the observed [O{\sc i}] line luminosities. In 
addition, we compute the emerging line profiles expected for 
either an outflowing disk or a Keplerian rotating disk, which can directly 
be compared to the observed profiles.
}
%Results
{
Both disk scenarios deliver reasonably good fits to the line luminosities and 
profiles of the [O{\sc i}] lines. Nevertheless, the Keplerian disk model seems
to be the more realistic one, because it also agrees with the kinematics 
derived from the large number of additional lines in the spectrum. As  
additional support for the presence of a high-density, gaseous disk, the 
spectrum shows two very intense and clearly double-peaked 
[Ca{\sc ii}] lines.
We discuss a possible disk-formation mechanism, and similarities 
between S\,65 and the group of Luminous Blue Variables.  
}
%Conclusions
{}

\keywords{
supergiants -- 
stars: winds, outflows -- 
stars: mass-loss --
circumstellar matter -- 
stars: individual: \object{LHA 115-S 65} 
                         }
\authorrunning{Kraus et al.}
\titlerunning{Neutral material around LHA\,115-S\,65}
\maketitle
%
%________________________________________________________________

\section{Introduction}

B[e] supergiants in the Large and Small Magellanic Clouds (LMC and SMC), even 
though studied in great detail, are still far from being understood. 
A non-spherically symmetric wind is suggested by, e.g., 
polarimetric observations (Magalh\~aes \cite{Magalhaes}; Magalh\~aes et al. 
\cite{Magalhaesetal}; Melgarejo et al. \cite{Melgarejo}), and 
a circumstellar disk seems to be confirmed by their strong infrared 
excess emission due to circumstellar dust (e.g., Zickgraf et al. 
\cite{Zickgraf86}) as well as by the emission of intense molecular bands like 
the CO first-overtone bands in the near 
infrared $K$-band spectra (McGregor et al. \cite{McGregor_a,McGregor_b,
McGregor89}; Morris et al. \cite{Morris}), and TiO bands at optical wavelengths
(Zickgraf et al. \cite{Zick89}). Based on the analysis of optical [O{\sc i}] 
lines, it has recently been suggested that the disks around B[e] supergiants 
are neutral in hydrogen right from (or at least very close to) the stellar 
surface (Kraus \& Borges Fernandes \cite{KB}; Kraus et al. \cite{Vlieland,
KBA}), so that molecules and dust are forming in the vicinity 
of the star. Indeed, for the LMC B[e] supergiant R\,126, Kastner et al. 
(\cite{Kastner}) found that the inner edge of the massive dusty disk must be 
located at $\sim 360\,R_*$, which is about three times closer to the star than 
the value of $\sim 1000\,R_*$ formerly suggested by Zickgraf et al. 
(\cite{Zickgraf85}). The much closer inner edge of the dust disk requests that
the [O{\sc i}] emission must originate from distances between the surface of
R\,126 and $\sim 360\,R_*$, which was confirmed by Kraus et al. (\cite{KBA}) 
considering a hydrogen neutral disk right from the stellar surface.

To guarantee that the disk material close to a luminous and hot supergiant star 
is neutral, the disk has to be massive (e.g., Kraus \& Lamers \cite{KL}). 
Because
the disks around supergiants cannot be pre-main sequence in origin, their 
formation must be linked to non-spherical (predominantely equatorial), 
high-density mass loss from the central star, either by some continuous steady 
material outflow, or by some 
mass-ejection event(s).

The formation of this anisotropic, high-density mass loss is assumed to be 
linked to rapid stellar rotation (e.g. Maeder \& Meynet \cite{MaMe00}; Kraus
\cite{Kraus06}). 
Especially rotation close to the critical velocity could trigger the (rotationally 
induced) bi-stability mechanism (Lamers \& Pauldrach \cite{LP}; Pelupessy et 
al. \cite{Pelupessy}; Cur\'{e} \cite{Cure04}; Cur\'{e} et al. \cite{Cure05}) as
a plausible disk-formation scenario. And indeed, for at least four members of 
the B[e] supergiant group, the projected rotation speed (i.e., $\varv \sin i$) 
could be determined, showing that two of them are rotating at a substantial 
fraction (i.e., at least 75\%) of their critical velocity. These are the SMC 
stars \object{LHA 115-S 23} discussed in detail by Kraus et al. 
(\cite{Kraus08}) and \object{LHA 115-S 65} (Zickgraf \cite{Zickgraf00}).
Whether all B[e] supergiants are rapidly rotating, is unknown however. The 
high-density wind observed around the other B[e] supergiants hides the 
central star, and the circumstellar material additionally contributes and 
pollutes the optical spectrum with a huge amount of emission lines. 
Consequently, the detection of uncontaminated photospheric lines, which would 
be appropriate for the determination of the projected stellar rotation 
velocity, is strongly hampered.
   
\begin{table}
  \begin{center}
  \caption{Parameters of S\,65. The rotation velocity was derived by 
Zickgraf (\cite{Zickgraf00}), the remaining parameters by Zickgraf et al. 
(\cite{Zickgraf86}).} 
  \label{tab1}
  \begin{tabular}{cccccc}\hline
  \hline
$T_{\rm eff}$ & $R_{*}$ & $\log L_*/L_{\odot}$ & $\varv\sin i$ & $E(B-V)$ & inclination \\
$ $ [K] &  [$R_{\odot}$] &   & [km\,s$^{-1}$] & &  \\
\hline
17\,000 & 81 & 5.7 & $\sim 150$ & $0.18\pm 0.02$ & $\pm$ edge-on \\
\hline
\end{tabular}
\end{center}
\end{table}

We study here the neutral material around the rapidly rotating SMC B[e] star 
LHA\,115-S\,65 (in short: S\,65). The star S\,65 is one of the four confirmed 
B[e] supergiants in the SMC. Its stellar parameters are summarized in 
Table\,\ref{tab1} and have been determined by Zickgraf et al. 
(\cite{Zickgraf86}) and Zickgraf (\cite{Zickgraf00}). Our investigation makes 
use of the three optical [O{\sc i}] emission lines detected in our spectra. We 
particularly aim to simultaneously reproduce the line luminosities and line 
profiles, to test the validity of any of the two most plausible B[e] supergiant 
disk scenarios: (i) an outflowing disk, which is neutral in hydrogen straight 
from the stellar surface, and (ii) a 
Keplerian rotating 
disk.

%__________________________________________________________________

\section{Observations and reductions}

We obtained high-resolution optical spectra using the Fiber-fed 
Extended Range Optical Spectrograph (FEROS), and also low-resolution optical 
spectra using the Boller \& Chivens spectrograph (B\&C).
                                                                                
The FEROS spectra were obtained in three different years. The oldest one is
from 1999 October 27, when the spectrograph was still attached to the 1.52-m 
telescope at the European Southern Observatory in La Silla (Chile), while the 
other spectra were taken on 2005 December 12, and 2007 October 3 and 4, after 
FEROS was moved to the 2.2-m telescope. FEROS is a bench-mounted Echelle 
spectrograph with fibers, where each one covers a sky area of 2$\arcsec$ of 
diameter. The wavelength coverage goes from 3600\,\AA \ to 9200\,\AA \ and the 
spectral resolution is $R = 55\,000$ (in the region around 6000\,\AA). The 
spectra were obtained with exposure times of 7200 (1999), 900 (2005), and 900 
(2007) seconds respectively. In 2005 and 2007, two exposures have been taken,
which were then added for a better S/N ratio. The S/N ratio in the 
5500\,\AA \ region is approximately 90 (1999), 40 (2005), and 65 (2007). 
We adopted the complete automatic on-line reduction.
                                                                                
\begin{figure*}[t!]
\begin{center}
\resizebox{\hsize}{!}{\includegraphics{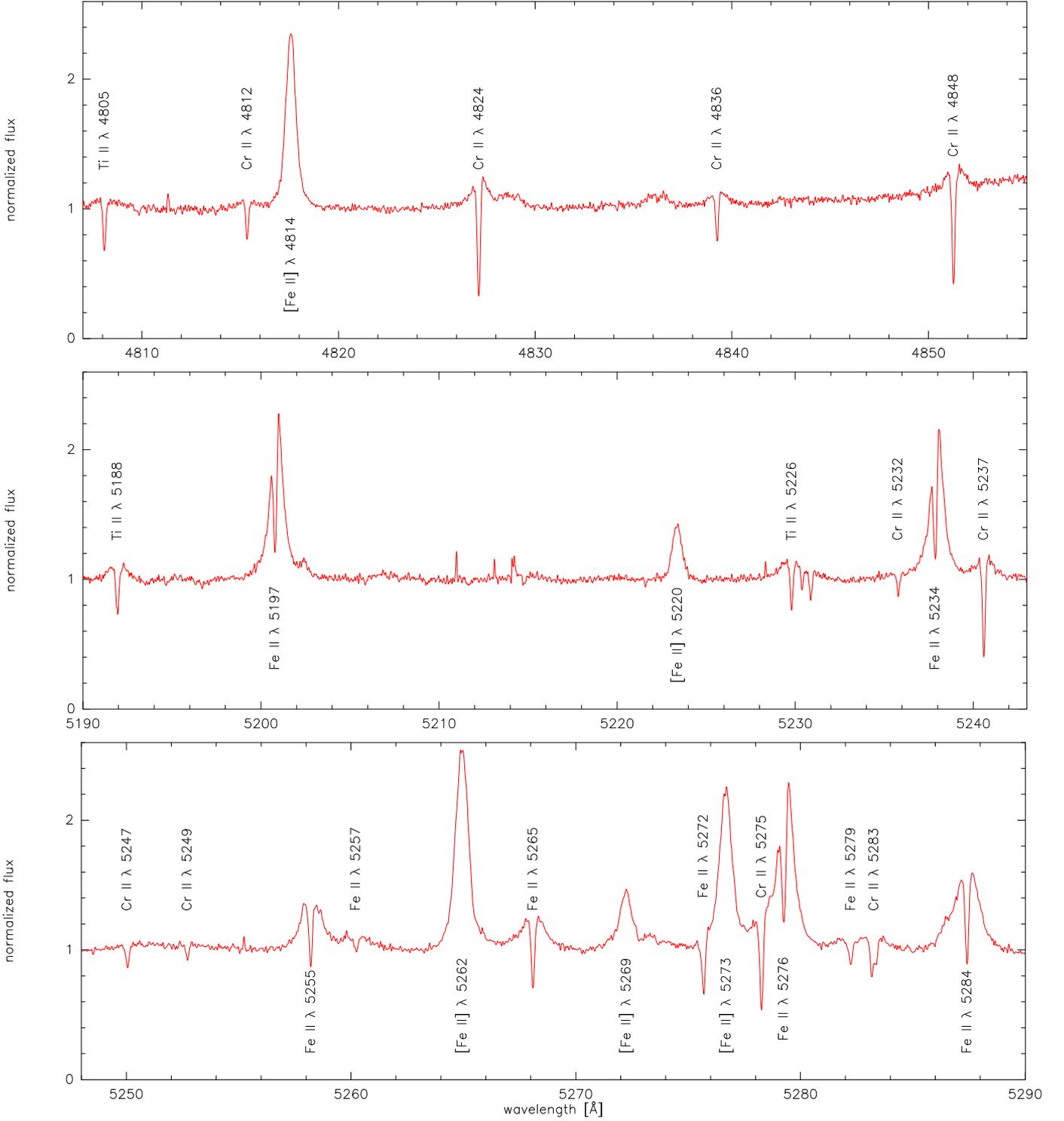}}
 \caption{Parts of our FEROS spectrum showing the two types of emission lines 
discussed in the text, i.e. the single-peaked emission lines of the 
forbidden transitions, and the permitted emission lines which all display a 
central or slightly blue-shifted sharp absorption component.}
 \label{spec}
 \end{center}
\end{figure*}

The low-resolution B\&C spectrum was taken on 1999 October 28, with an 
exposure time of 900 seconds and a slit width of 4$\arcsec$. The instrumental 
setup employed provides a resolution of $\sim 4.6$\,\AA \ in the range of 
3800-8700\,\AA. The S/N ratio, in the 5500\,\AA \ continuum region, is 
aproximately 60. The B\&C spectrum was reduced with standard 
IRAF\footnote{IRAF is distributed by the National Optical Astronomy 
Observatories, which are operated by the Association of Universities for 
Research in Astronomy, Inc., under cooperative agreement with the National
Science Foundation.} tasks, such 
as bias subtraction, flat-field normalization, and wavelength calibration. The 
absolute flux calibration was made using spectrophotometric standard stars 
cited in Hamuy et al. (\cite{Hamuy}).
                                                                                
%__________________________________________________ 

\section{Results}

\subsection{Spectral characteristics}\label{spec_char}

A first detailed list of identified lines in the optical spectral range of 
S\,65 has been compiled by Muratorio (\cite{Muratorio}). Later on, Zickgraf 
et al. (\cite{Zickgraf86}) re-observed S\,65 at higher spectral resolution and 
with higher signal-to-noise ratio, and described their optical spectra in great 
detail. They found that the line spectrum showed many features in common with 
shell stars. These are the Balmer lines, which display a P\,Cygni profile of 
Beals' type III with a shell absorption reaching in most cases below the 
continuum, the Fe{\sc ii} emission lines with a central absorption core, and 
many additional singly ionized metal lines (like those from Cr{\sc ii} and 
Ti{\sc ii}, see Fig.\,\ref{spec}) showing very narrow central absorption 
components which are nearly unshifted with respect to the emission component. 
For the typical B[e] supergiants, these well pronounced shell 
lines were thus interpreted by Zickgraf et al. (\cite{Zickgraf86}) as clear 
signatures for an edge-on orientation of the S\,65 system. Further support for 
an edge-on orientation came from optical polarization measurements performed by 
Magalh\~aes (\cite{Magalhaes}) and Melgarejo et al. (\cite{Melgarejo}).
Follow-up observations with FEROS (Stahl \cite{Stahl}) showed no 
variability in the spectral lines of S\,65, when compared to the spectra of
earlier investigations. Indeed, comparing our own three spectra that cover a
period of eight years to those of Zickgraf et al. (\cite{Zickgraf86}), as well as 
to the FEROS spectrum of Stahl (\cite{Stahl}) that was obtained one year prior 
to our first spectrum, we could not find any significant difference in the 
individual line profiles, indicating that S\,65 is not undergoing any severe 
variations. Nevertheless, before going into any modeling details, we will first
provide a short description of the elements seen in our spectra and their 
line profiles. Due to the good quality of the spectra taken in 2007, we 
measured all the lines reported in this paper from these spectra. The spectra 
taken in 1999 and 2005 are used for a qualitative comparison.

\begin{figure}[t!]
\begin{center}
\resizebox{\hsize}{!}{\includegraphics{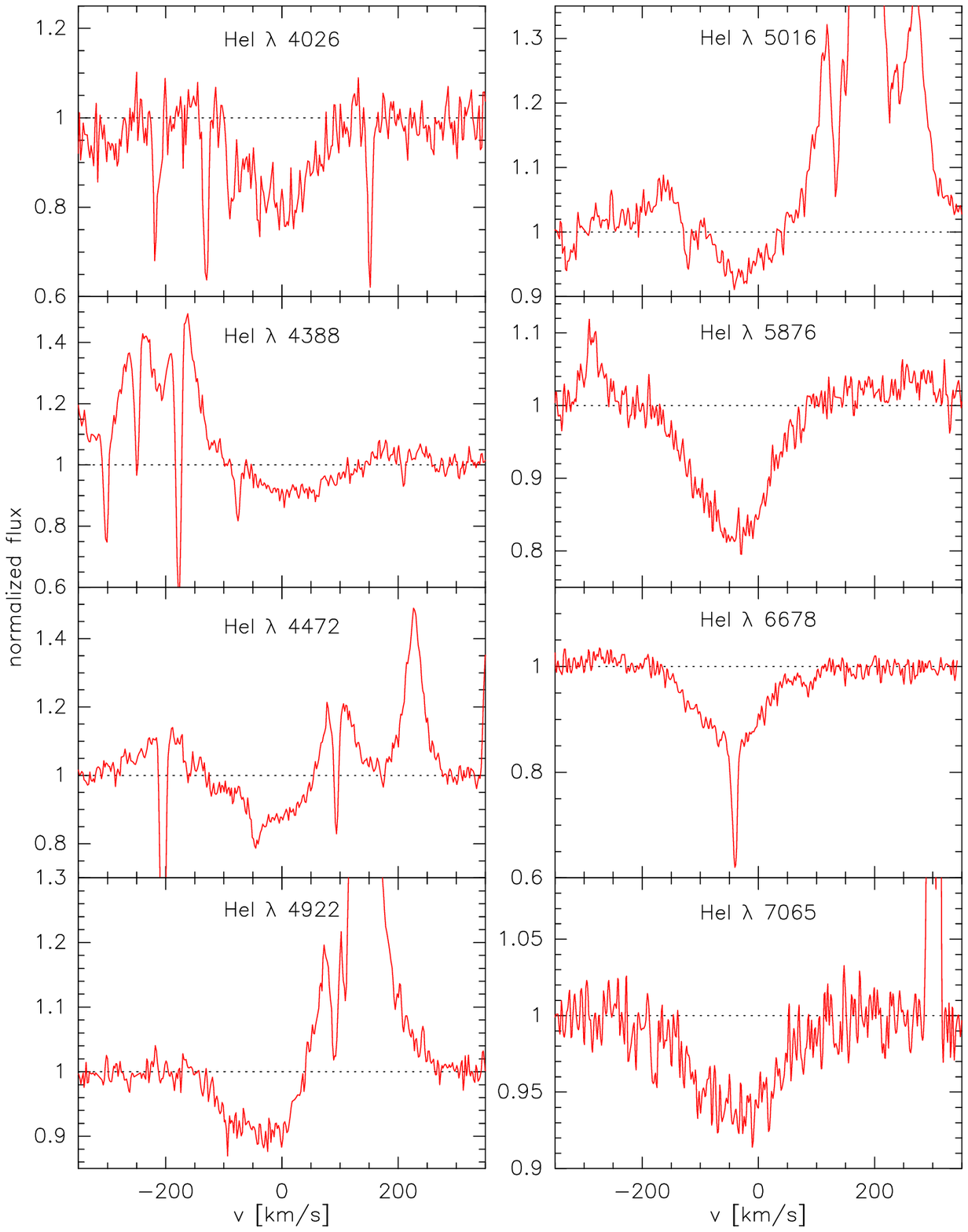}}
 \caption{ Photospheric He{\sc i} absorption lines. The lines 
are centered on the mean systemic velocity, which was determined as described 
in Sect.\,\ref{sys_rot}.}
 \label{He_pap}
 \end{center}
\end{figure}

\begin{figure}[t!]
\begin{center}
\resizebox{\hsize}{!}{\includegraphics{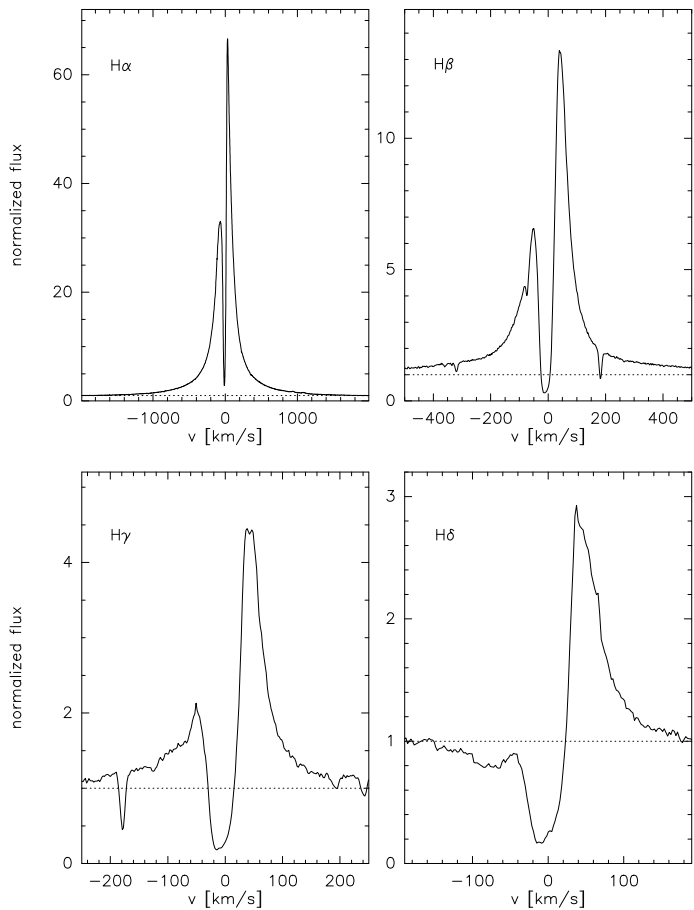}}
 \caption{Balmer line profiles showing a variation in shape. 
The lines are centered on the mean systemic velocity of S\,65, 
which was determined as described in Sect.\,\ref{sys_rot}.} 
 \label{balm}
 \end{center}
\end{figure}

\begin{figure}[t!]
\begin{center}
\resizebox{\hsize}{!}{\includegraphics{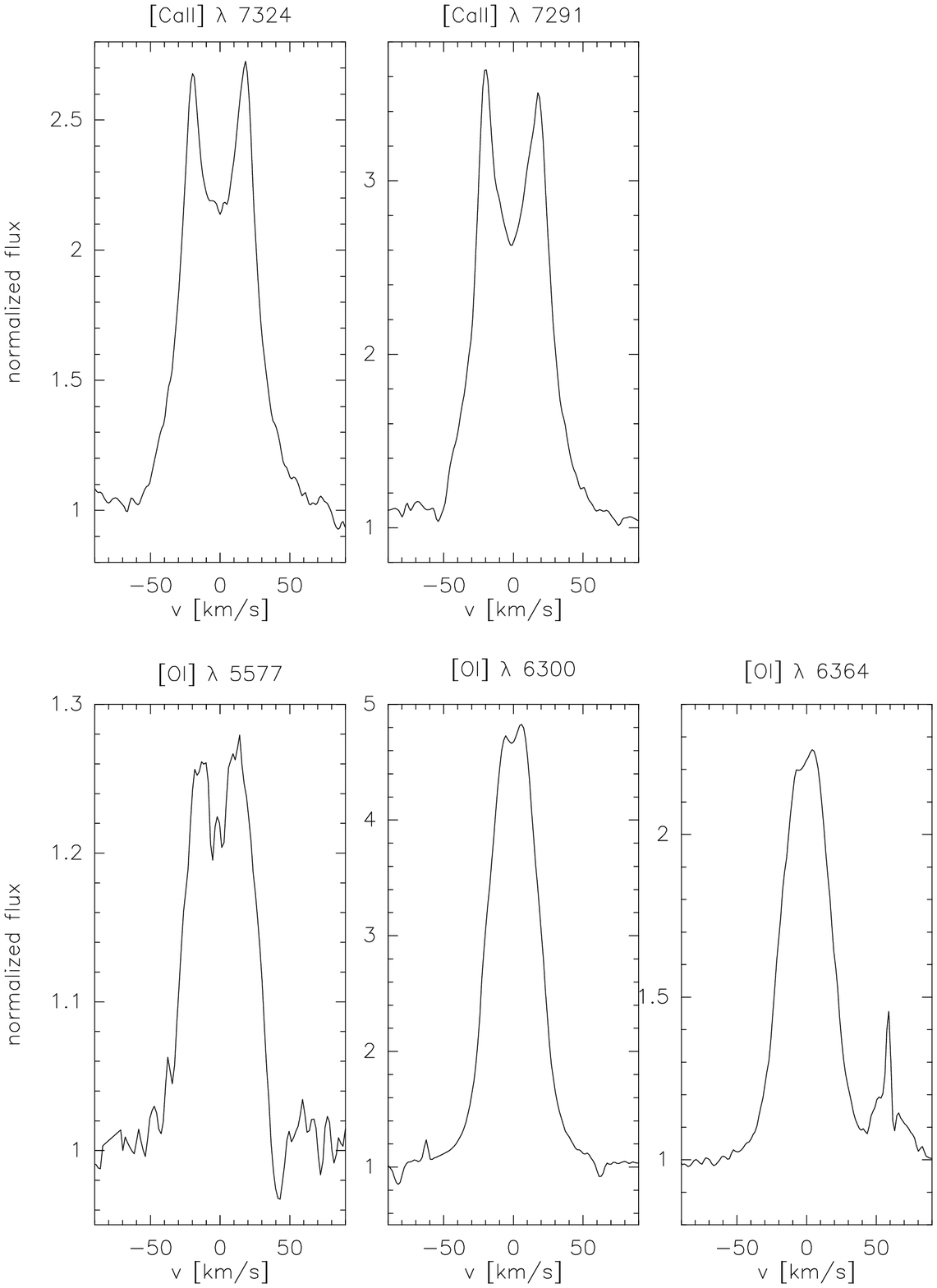}}
 \caption{Line profiles of the [Ca{\sc ii}] (top) and [O{\sc i}] lines 
(bottom). The lines are centered on the mean systemic velocity of S\,65,  
which was determined as described in Sect.\,\ref{sys_rot}.}
 \label{forb}
 \end{center}
\end{figure}

Figure\,\ref{spec} covers some small portions of the optical spectrum of S\,65. 
The numerous lines seen in these short wavelength ranges represent the richness 
in emission lines of the complete spectrum.
A detailed investigation revealed only a few very weak 
photospheric absorption lines. Ten lines are from He{\sc i}, which are all
at least partly contaminated by or blended with circumstellar emission lines.
This is shown for eight of them in Fig.\,\ref{He_pap}.
We also tentatively detected two lines from N{\sc ii}, $\lambda\lambda$ 3995 
and 5679, and possibly one line from O{\sc ii} at $\lambda$ 3954. 

The Balmer lines are present from H$\alpha$ up to H15. Figure\,\ref{balm} 
shows the line profiles of H$\alpha$ to H$\delta$. The lowest Balmer lines 
display double-peaked profiles with a so-called violet-to-red ratio $V/R 
< 1$ and a broad central absorption that 
extends below the continuum (except for H$\alpha$, see Fig.\,\ref{balm}). These
line profiles are well-known from classical Be stars (or Be shell
stars, e.g., Hanuschik \cite{Hanuschik}) viewed edge-on.
From H$\delta$ on the shape of the 
lines turn into a P\,Cygni profile. 
The strength of the emission component thereby quickly decreases with higher 
quantum numbers, until for Balmer lines higher than H13 only the absorption 
component remains. The absorption thereby consists of the P\,Cygni part and a 
shallow, but rather broad photospheric part, as can be seen already for the 
H$\delta$ line in Fig.\,\ref{balm}. Also present are lines of the higher 
Paschen series, i.e., Pa10 up to Pa24, displaying P\,Cygni-type profiles. 
For the Paschen lines the absorption component also consists of a rather weak 
P\,Cygni absorption component, extending into the very shallow but much broader 
photospheric component. 

The huge number of additional lines seen in our spectra can basically be 
divided into two groups: single-peaked emission lines, and emission lines 
with a strong but very narrow, almost central absorption component reaching in 
most cases far below the continuum. The two types of profiles are nicely seen
in the parts of the spectrum shown in Fig.\,\ref{spec}.

All single-peaked emission lines are found to come from forbidden 
transitions. Most of the lines belong to [Fe{\sc ii}], but a few lines of 
[Cr{\sc ii}] and [Ni{\sc ii}], as well as one line from [S{\sc ii}] are also 
visible. Additional lines (but blended or very weak) from [Ti{\sc i}], 
[Mg{\sc i}], [Mg{\sc ii}], and [N{\sc ii}] might be present. We also found a 
few very weak emission lines from permitted transitions; they belong to 
C{\sc i} and N{\sc i}. 

The group of emission lines with the strong central absorption contains by far 
the largest number of lines. We could identify more than 230 lines, most of 
them belonging to transitions of Fe{\sc ii}, Cr{\sc ii}, and Ti{\sc ii}.
But few lines from other elements like Mn{\sc ii}, Mg{\sc i}, Ca{\sc ii}, 
Si{\sc ii}, and Fe{\sc i} were also found. The strong, sharp, and almost 
central absorption components indicate high-density absorbing material along 
the line of sight with a rather small radial velocity component. As mentioned
earlier, the line shapes recall those seen in Be shell stars or 
Be disks viewed edge-on. It is therefore reasonable to assume that S\,65 is 
surrounded by a high-density disk or ring structure viewed edge-on, as 
suggested by Zickgraf et al. (\cite{Zickgraf86}).

Interestingly, the spectrum shows prominent emission of the [O{\sc i}] lines as 
well as of the [Ca{\sc ii}] lines. However, their line profiles do not fit into
the groups described above. Instead, their profiles show a clear double-peaked
structure (see Fig.\,\ref{forb}). This type of profile is usually interpreted 
with Keplerian rotation in a circumstellar disk.

\subsection{Systemic and projected rotational velocity of S\,65}\label{sys_rot}

\begin{figure}[t!]
\begin{center}
\resizebox{\hsize}{!}{\includegraphics{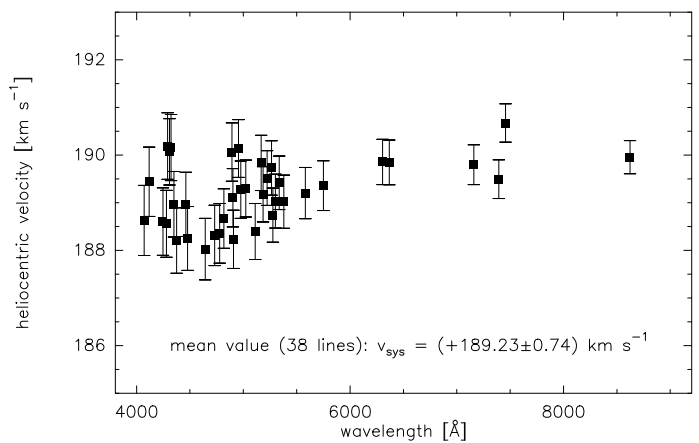}}
 \caption{Radial (heliocentric) velocity of S\,65 derived from the
strong and symmetric non-blended single-peaked forbidden emission lines.}
 \label{vrad}
 \end{center}
\end{figure}

The strong pollution of the photospheric He{\sc i} absorption lines by 
adjacent emission from the circumstellar material (Fig.\,\ref{He_pap})
makes it impossible to accurately determine their line centers. Therefore we 
need to rely on the emission lines to derive the systemic velocity.
As mentioned in the previous section, the permitted lines all 
show a sharp central absorption. The slight asymmetry seen in most of these 
line profiles in form of a slightly stronger red emission compared to the blue 
emission peak hampers a proper determination of their line centers. In 
addition, this asymmetry indicates that the central absorption components seem
to be in many cases slightly blue-shifted. Hence they do not
mirror the systemic velocity of the system, but only the dynamics along the
line-of-sight through the high-density disk (see Sect.\,\ref{discussion}).
The only reasonable set of lines for a proper systemic velocity determination 
is thus provided by the optically thin, single-peaked and symmetric profiles of 
the forbidden emission lines. From the whole sample of forbidden lines 
detected, we restricted our selection to the strongest and unblended ones and 
measured their central wavelengths by fitting Gaussian profiles using IRAF. 
From the difference between the measured and the 
laboratory wavelengths we computed the radial velocity, which is plotted in
Fig.\,\ref{vrad}. The mean radial (heliocentric) velocity we found is
$\varv = (+189.23\pm 0.74)$\,km\,s$^{-1}$, which can be interpreted as the 
systemic velocity of S\,65. Our mean value agrees very well with the 
value of ($+191\pm 4$)\,km\,s$^{-1}$ derived by Zickgraf et al. 
(\cite{Zickgraf86}), and also fairly well with the 
values of $+184$\,km\,s$^{-1}$, and ($+193\pm 
6$)\,km\,s$^{-1}$ obtained in earlier investigations by 
Feast et al. (\cite{Feast}) and Maurice (\cite{Maurice}), 
respectively, when we account for the lower resolutions and lower S/N values 
of those older spectra. Therefore we consider our value as reasonably good
and use it throughout the paper to correct the measured velocities with 
respect to the systemic velocity, as done, e.g., for the line profile plots in
Figs.\,\ref{He_pap} to \ref{forb}.

We emphasize that the He{\sc i} absorption lines cannot be 
formed in the (polar) wind of S\,65, because the effective temperature of the 
star (and hence the wind temperature) is too low to keep helium ionized beyond 
the photoshpere. Therefore they can be considered as of pure photospheric
origin.
The width of these He{\sc i} lines and also of the (shallow) photospheric
absorption components of the Balmer and Paschen lines cannot be caused by 
broadening mechanisms like macroturbulence and pressure broadening alone. These
two effects in the atmospheres of B-type supergiants typically add up to a few 
tens of km\,s$^{-1}$, only. Instead, the broad and roundish shape of the lines 
suggests that they are predominantely broadened by the possibly high rotation 
of the star, as suggested by Zickgraf (\cite{Zickgraf00}). 

\begin{figure}[t!]
\begin{center}
\resizebox{\hsize}{!}{\includegraphics{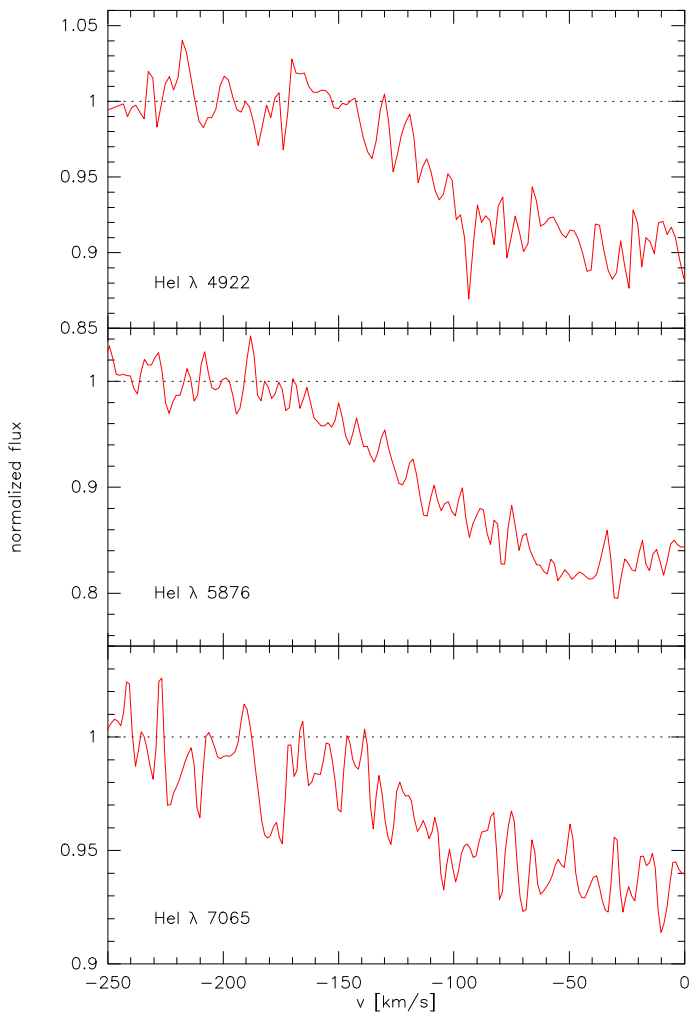}}
 \caption{Uncontaminated blue parts of three He{\sc i} lines from
Fig.\,\ref{He_pap}.}
 \label{He_rot}
 \end{center}
\end{figure}

Our derived systemic velocity allows us to determine the proper line 
centers of the He{\sc i} lines. Inspection of these lines (see 
Fig.\,\ref{He_pap}) shows that the blue parts of at least three lines 
(He{\sc i} $\lambda\lambda$ 4922, 5876, and 7065) seem to be uncontaminated, 
nicely displaying the roundish profile shape as expected in rotationally 
broadened photospheric absorption lines. The blue half of these three He{\sc i} 
lines are shown in Fig.\,\ref{He_rot}. Their wings extend to Doppler 
velocities of about 150--180\,km\,s$^{-1}$. A comparable Doppler
velocity is also seen in the shallow photospheric absorption components of the 
hydrogen lines (see, e.g., the line profile of H$\delta$ in Fig.\,\ref{balm}).

We measured the half-width at half-maximum ($HWHM$) values of these three 
He{\sc i} lines in Fig.\,\ref{He_rot}. From those values, the projected 
rotation velocity, $\varv\sin i$, can 
be computed under the assumption of a stellar rotation profile as derived, 
e.g., by Gray (\cite{Gray}) and with a limb-darkening coefficient of 0.6. The 
measured $HWHM$ and the derived $\varv\sin i$ values are listed in 
Table\,\ref{rotvelo}. Compared to the value of $\sim 150$\,km\,s$^{-1}$ 
obtained by Zickgraf (\cite{Zickgraf00}), our values excellently 
agree. Given that the He{\sc i} $\lambda$5876 line is the strongest in our 
sample, we consider our value of $(155.2\pm 6.0)$\,km\,s$^{-1}$ derived from 
that line as a reasonably good value for the projected rotation velocity of 
S\,65. 

\begin{table}
  \begin{center}
  \caption{Measured $HWHM$ and derived $\varv\sin i$ values of the three
He{\sc i} lines in Fig.\,\ref{He_rot}.}
  \label{rotvelo}
  \begin{tabular}{lcc}\hline
  \hline
Line & $HWHM$ & $\varv\sin i$ \\
     &  [\AA]          & [km\,s$^{-1}$]        \\
\hline
He{\sc i} 4922 & $1.81\pm 0.10$ & $141.8\pm 7.4$ \\
He{\sc i} 5876 & $2.37\pm 0.09$ & $155.2\pm 6.0$ \\
He{\sc i} 7065 & $2.91\pm 0.14$ & $158.1\pm 7.5$ \\
\hline
\end{tabular}
\end{center}
\end{table}
 
We note however that this value might be un underestimation of the real
projected stellar rotation velocity. Expressed in terms of the critical 
rotation velocity, the derived value of $\sim 155$\,km\,s$^{-1}$ corresponds 
to $\varv\sin i/\varv_{\rm crit}\simeq 0.75$. For stars rotating so rapidly
Townsend et al. (\cite{Townsend}) have shown that the $HWHM$ method is not 
reliable anymore to derive the correct projected stellar rotation velocity,
because the effects of gravitational darkening influence the photospheric
absorption lines in a way that their line width does not increase anymore for
$\varv\sin i/\varv_{\rm crit} > 0.75$. This means that even if the star would 
rotate at a velocity close to the critical one, the measured $HWHM$ would only 
imply a value of $\varv\sin i/ \varv_{\rm crit}\simeq 0.75$. Our value
can thus only be considered as a lower limit to the real projected stellar
rotation of S\,65.

\subsection{The [O{\sc i}] lines}\label{oi_theorie}

In the remaining part of this section we concentrate on the [O{\sc i}] 
emission lines arising in the circumstellar material of S\,65, aiming to find 
a resonable scenario to explain the observed [O{\sc i}] line luminosities and 
profiles.
                                                                                
The FEROS spectrum displays three [O{\sc i}] lines (see Fig.\,\ref{forb}). They 
originate from transitions within the five lowest energy levels of neutral 
oxygen. Based on the flux-calibrated (but low-resolution) B\&C spectrum, we
can derive the line luminosities of the [O{\sc i}] lines. Zickgraf et al. 
(\cite{Zickgraf86}) determined an interstellar extinction towards S\,65 of 
$E(B-V) = 0.18\pm 0.02$ (see Table\,\ref{tab1}). We used this value and applied 
the interstellar extinction curve of the SMC of Pr\'{e}vot et al. 
(\cite{Prevot}) to deredden the B\&C spectrum. With the calibrated dereddened 
continuum and a distance modulus of $\mu = 18.93\pm 0.02$ for the SMC (Keller 
\& Wood \cite{Keller}), we then calculated the observed luminosities in the 
[O{\sc i}] lines (see Table\,\ref{peaks}) from the measured equivalent widths 
in the FEROS spectrum. 
The errors in line luminosities are about 10\%, following from the 
uncertainties in equivalent width measurements, interstellar extinction, and 
distance.

The major advantage in using forbidden emission lines is that because they are 
optically thin, their line profiles contain the full velocity information of 
their formation region. The observed line profiles of the [O{\sc i}] 
$\lambda$\,6300 and 6364 lines are more or less identical. This is clear, 
because both lines arise from the same upper level, which means that they are 
formed in the same region in terms of density and temperature. While both lines 
show only a slight indication of a double-peaked profile, this double-peak 
structure is more pronounced in the [O{\sc i}] $\lambda$\,5577 line, and the 
measured peak velocities and $FWHM$ values of all three lines are listed in 
Table\,\ref{peaks}. This latter line arises from a higher energy level. 
Because the levels from which the
forbidden lines originate are purely collisionally excited, the [O{\sc i}]
$\lambda$\,5577 line must be formed in regions of a higher density and/or
higher temperature. This could indicate that the $\lambda$\,5577 line 
originates from regions closer to the star. In combination with the broader
$FWHM$ as well as the broader peak seperation, which mirror a higher velocity 
within the $\lambda$\,5577 line-forming region, this qualitative analysis
might already hint at a Keplerian disk rather than a pure outflowing
disk.

However, based purely on the appearance of a double-peaked structure of the 
line profiles, it is not possible to discriminate whether the [O{\sc i}] 
line-forming region is located in a Keplerian rotating disk or in an outflowing 
disk, because both scenarios, when viewed edge-on, result in identical, 
double-peaked line profiles. Reliable conclusions about the real kinematical 
nature of the [O{\sc i}] line-forming region can thus only be drawn from a 
proper and simultaneous modeling of both line profiles, from which the 
kinematical information is taken and the line luminosities, delivering the 
information about the density and temperature. This is done below for the two 
competing scenarios: the outflowing and the Keplerian rotating disk model.

\begin{table}[t!]
  \begin{center}
  \caption{Velocities obtained from the peak separation and $FWHM$ values and 
extracted line luminosities of the [O{\sc i}] lines. The errors in the line 
luminosities are on the order of 10\%.}
  \label{peaks}
  \begin{tabular}{cccc}\hline
  \hline
$\lambda$ & $\varv_{\rm peaks}$ & $\varv_{FWHM}$ & $L_{\rm line}$ \\
$[$\AA$]$ & [km\,s$^{-1}$] & [km\,s$^{-1}$] &
[erg\,s$^{-1}$] \\
\hline
5577 & $44\pm 2$ & $56\pm 2$ & $8.28\times 10^{33}$\\
6300 & $24\pm 2$ & $43\pm 2$ & $8.46\times 10^{34}$\\
6364 & $24\pm 2$ & $42\pm 2$ & $2.93\times 10^{34}$\\
\hline
\end{tabular}
\end{center}
\end{table}

\subsubsection{Outflowing disk model}\label{outflow}

Not much is known in the literature concerning the structure and shape of the 
disks around B[e] supergiants, and there is an ongoing debate whether they
are Keplerian rotating, or outflowing in nature (see, e.g., Porter \cite{John};
Kraus et al. \cite{KBA}). From a statistical point of view (based on low 
number statistics though), it seems that the disks must be rather thick, with 
an opening angle of about $\alpha \simeq 20\degr$ (Zickgraf 
\cite{Zickgraf89}), which is 4--5 times larger than the opening angle of 
classical Be stars (see, e.g., Porter \& Rivinius \cite{Porter}).

Another key question is related
to the ionization structure in the disk. Oxygen has about the same ionization 
potential as hydrogen. The detection of emission lines from neutral oxygen thus 
means that these lines must be generated within a region in which hydrogen is 
predominantly neutral as well. Free electrons are (besides the less 
efficient, but nevertheless important, neutral hydrogen atoms) the main 
collision partners to excite the lowest energy levels in neutral oxygen, 
from which 
the forbidden emission lines arise. We follow the approach of Kraus et al.
(\cite{KBA}) and introduce the parameter $q_{\rm e}$ as the ionization 
fraction, defined by $n_{\rm e} = q_{\rm e} n_{\rm H}$, with $q_{\rm e} = 
q_{\rm Metals} + q_{\rm H^{+}}$, and the electron and hydrogen particle 
densities, $n_{\rm e}$ and $n_{\rm H}$, respectively. Assuming that all 
elements with an ionization potential lower than that of hydrogen and oxygen 
are fully ionized delivers an upper limit of $q_{\rm Metals}<9.5\times 10^{-5}$ 
For this, we used an SMC metallicity of 20\% solar (see, e.g., Dufton et al. 
\cite{Dufton}), and solar abundance values for the metals from Grevesse \&
Sauval (\cite{Grevesse}). This very low value obtained from the metals means 
that if hydrogen is ionized by only 1\%, it still provides the dominant 
quantity of free electrons. Because we do not know the amounts of ionized 
hydrogen and ionized metals within the [O{\sc i}] line-forming region, we keep 
$q_{\rm e}$ as a free parameter at first.

An additional important parameter is the electron temperature in the [O{\sc i}] 
line-forming region. It should be lower than the hydrogen ionization temperature,
and higher than the molecule dissociation threshold, below which atomic oxygen 
is bound into CO molecules. While the former restriction defines the maximum 
temperature at about 10\,000\,K, the latter corresponds to a minimum 
temperature of roughly 5000\,K, which is the dissociation temperature of CO. 

To calculate the population of the five lowest energy levels in atomic oxygen,
we need to take into account collisions from both free electrons and neutral
hydrogen atoms. The collision parameters for excitation by neutral hydrogen 
atoms are taken from St\"{o}rzer \& Hollenbach (\cite{Stoerzer}), those for 
excitation by free electrons from Mendoza (\cite{Mendoza}), and atomic 
parameters are from Wiese et al. (\cite{Wiese}) and Kafatos \& Lynch 
(\cite{Kafatos}). We calculate the level population by solving the statistical 
equilibrium equations in a 5-level atom. Forbidden emission lines are optically
thin so that no complicated radiation transfer needs to be solved. Instead,
the emissivity at any location $r$ in the disk is simply given by
$j_{n,m}(r) = h\nu_{n,m}n_{n}(r)A_{n,m}$ where $n_{n}(r)$ is the level
population of the upper level, and $A_{n,m}$ is the Einstein coefficient of
spontaneous emission. The line luminosities finally follow from integration of 
the emissivities over the emitting disk volume.

Next, we can restrict the valid ranges in electron temperature and 
ionization fraction, $q_{\rm e}$, within the [O{\sc i}] line-forming regions in 
the disk of S\,65. Because two of the observed [O{\sc i}] lines ($\lambda 6300$ 
and $\lambda 6364$) originate from the same upper level,
their line ratio is determined by pure quantum mechanics and does not depend on 
density and temperature. The $\lambda 5577$ line, however, originates from a 
higher energy level. A sufficiently strong population of this level thus 
severely depends on the density of the collision partners as well as on 
temperature. Any ratio with this line is therefore an ideal tracer for the 
density and temperature, and we concentrate on the $\lambda 6300/\lambda 5577$ 
line ratio below to constrain these two parameters. 

\begin{figure}[t!]
\begin{center}
\resizebox{\hsize}{!}{\includegraphics{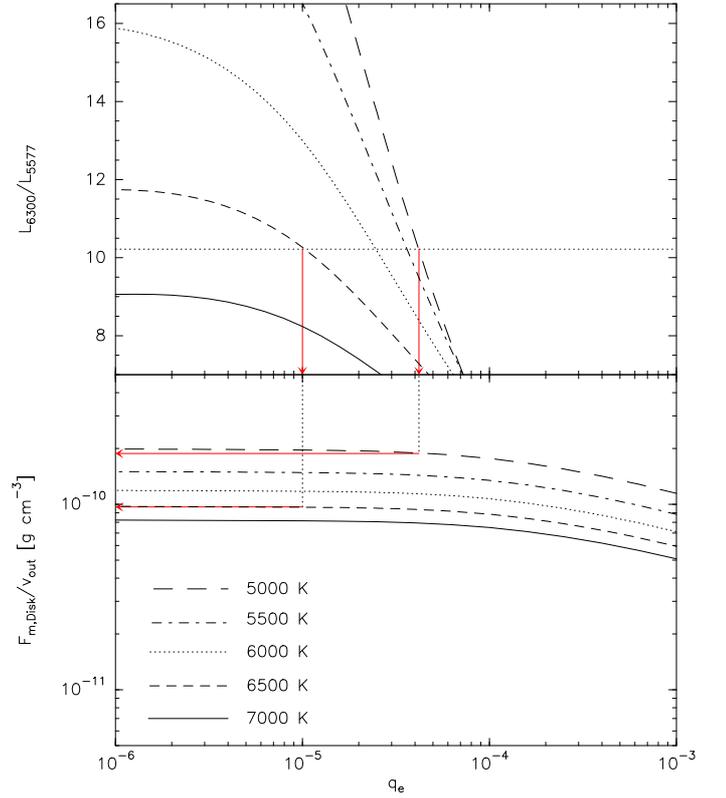}}
 \caption{
{\it Top panel:} Theoretical line ratios as function of $q_{\rm e}$ at 
fixed temperature. The intersection with the observed value (dotted horizontal 
line) defines the valid ($q_{\rm e}, T_{\rm e}$) range (arrows).
{\it Bottom panel:} Density parameter as function of $q_{\rm e}$ at fixed 
temperature. The valid ($q_{\rm e}, T_{\rm e}$) range (dotted vertical lines) 
delivers the range in density parameters (arrows). 
}
 \label{param}
 \end{center}
\end{figure}

In the top panel of Fig.\,\ref{param} we show the $\lambda 6300/\lambda 5577$ 
line ratio as a function of the ionization fraction, $q_{\rm e}$, and for 
different (constant) values of the electron temperature. The calculations  
performed with the outflowing disk model request the knowledge of the mass 
flux. For simplicity, we assume a latitude independent mass flux within the 
disk-forming wind region, i.e., $F_{\rm m}(\theta) = F_{\rm m, Disk} = {\rm 
const.}$ for all latitudes in the range $[90\degr - (\alpha/2)] < \theta < 
[90\degr + (\alpha/2)]$. We also assume the wind to have a constant 
ouflow velocity, $\varv_{\rm out}$, delivering the density parameter, $F_{\rm 
m, Disk}/\varv_{\rm out}$, as one free input parameter. With this density 
parameter we can calculate the hydrogen density at the inner edge of the disk 
(here: at the stellar surface) and consequently also the radial density 
distribution $n_{\rm H}(r) \sim (F_{\rm m, Disk}/\varv_{\rm out})\,r^{-2}$ 
resulting from the application of the equation of mass continuity. Due to the 
radially decreasing density, the [O{\sc i}] line luminosities saturate at a 
certain distance. This saturation value should agree with the observed line 
luminosities. We thus varied the density parameter, $F_{\rm m, Disk}/\varv_{\rm 
out}$, for a fixed pair of ($T_{\rm e}, q_{\rm e}$) values, until the observed 
$\lambda 6300$ line luminosity was reproduced. The line ratio with the $\lambda 
5577$ line resulting from these calculations defines one single point in the 
diagram shown in the top panel of Fig.\,\ref{param}. This procedure was 
repeated until a sufficiently large coverage of the ($T_{\rm e}, q_{\rm e}$) 
parameter space was obtained. We then compared the theoretical results to the 
observed line ratio, which is shown as the dotted line. Based on the discussion 
above concerning the valid range in temperatures, the curve for $T_{\rm e} = 
5000$\,K can be considered as the lower temperature limit, resulting in a {\it 
strict upper limit} for the ionization fraction in the disk of S\,65 of $q_{\rm 
e, max}=4.2\times 10^{-5}$. This means that the [O{\sc i}] line-forming region 
must indeed be neutral in hydrogen, with $q_{{\rm H}^{+}} \la q_{\rm e, max} = 
4.2\times 10^{-5}$. For a temperature of 7000\,K, the observed line ratio 
cannot be reproduced. The maximum temperature in the [O{\sc i}] line-forming 
region must thus lie between 6500\,K and 7000\,K. Since the plot in the top 
panel of Fig.\,\ref{param} indicates that a higher temperature would imply a 
lower degree of ionization, it is reasonable to assume that the most plausible 
range in temperature is $T_{\rm e}\simeq 5500\ldots 6500$\,K, delivering a disk 
ionization fraction of $q_{\rm e}\simeq (1\ldots 4)\times 10^{-5}$.

In the bottom panel of Fig.\,\ref{param} we plot the density parameter, used 
for the line luminosity fit of the $\lambda 6300$ line and the resulting line 
ratio as shown in the top panel, as a function of $q_{\rm e}$ and for the 
different temperature values. In the part of the diagram where the ionization 
fraction is high, the density parameter shows (for a given temperature) a power-law 
dependence with $q_{\rm e}$. This means that for decreasing ionization fraction 
the input density has to be increased to compensate for the otherwise 
decreasing number of free electrons. On the other hand, below an ionization 
fraction of a few times $10^{-5}$, the input density parameters stagnate. This 
means that for ionization fractions $q_{\rm e}\la 10^{-4}$ collisions with 
neutral H atoms start to become important. For further decreasing 
ionization fraction they even dominate the level population process,
because at very low ionization the neutral particles are finally 
numerous enough for efficient collisional level population. From the narrow 
range of valid temperatures and ionization fraction values, the range in 
density parameter is confined to $F_{\rm m, Disk}/\varv_{\rm out}\simeq (1
\ldots 2)\times 10^{-10}$\,g\,cm$^{-3}$.

\begin{figure}[t!]
\begin{center}
\resizebox{\hsize}{!}{\includegraphics{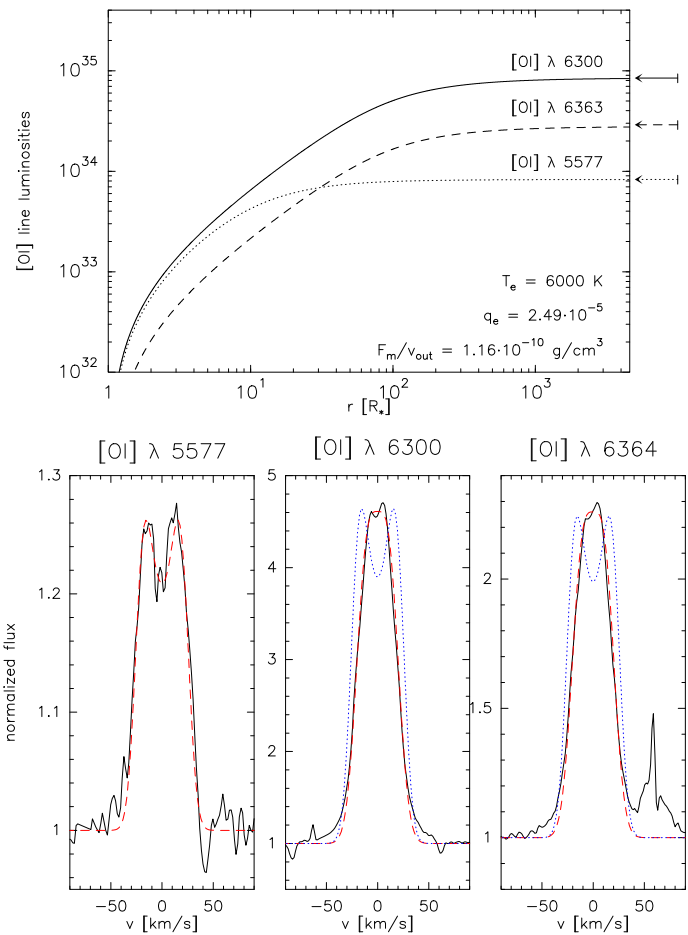}}
\caption{{\sl Top panel:} Predicted [O{\sc i}] line luminosities from
within a radius $r$ in the 
outflowing disk scenario. The arrows to the right indicate the observed values.
{\sl Bottom panel:} Fits to the line profiles. The dotted profiles fitted to 
the $\lambda 6300$ and $\lambda 6364$ lines are obtained for the same constant 
outflow velocity as the $\lambda 5577$ line; the dashed lines in all profiles 
show our best fit.}
\label{wind}
\end{center}
\end{figure}

These calculations, even though they are performed under several simplifying
assumptions, show that the [O{\sc i}] line-forming region is well constrained in
temperature and density. And the increase of the individual line luminosities
with distance from the star and their saturation is shown for all three [O{\sc 
i}] lines in the upper panel of Fig.\,\ref{wind}. These computations 
exemplify the case of a constant temperature of 6000\,K and 
resulting values of $2.49\times 10^{-5}$ and $1.16\times10^{-10}$\,g\,cm$^{-3}$ 
for the ionization fraction and density parameter, respectively. The arrows to 
the right indicate the observed line luminosity values with their errors 
indicated by the vertical bars.

The luminosity saturation of the $\lambda 5577$ line occurs at a distance of 
about $20-30\,R_*$, and therefore rather close to the star, while saturation in 
the luminosities of the other two lines happens only at distances beyond $\sim 
400\,R_*$, i.e., at least 10 times further out and therefore in regions in 
which the density is at least 100 times lower. To check the influence of the 
disk temperature on these results we also calculated the line luminosities for 
a disk with 6500\,K and 5500\,K. A lower (higher) temperature requires slightly 
higher (lower) values for the ionization fraction and density parameter (see 
Fig.\,\ref{param}).
The corresponding curves for the increases in line luminosities are, however, 
hardly affected, and the saturations of the [O{\sc i}] lines happen at the 
same distances. What remains to check is whether this outflowing disk scenario 
is a reasonable model to explain the observed line profiles.

Because the lines are formed in different disk regions, their profiles mirror 
the kinematics at these different distances from the star. The double-peaked 
profile of the $\lambda 5577$ line might be interpreted with an outflow 
velocity of about $\varv_{\rm out}\simeq 22$\,km\,s$^{-1}$, obtained from the 
separation between the blue edge of the blue and the red edge of the red peak 
(Table\,\ref{peaks}). The broader wings of the line require an additional 
Gaussian-shaped velocity component, i.e., additional to the thermal velocity 
plus spectral resolution, which result in a Gaussian velocity component of 
about 6\,km\,s$^{-1}$. This additional Gaussian velocity could be ascribed to 
some turbulent motion of the gas and is found to be on the order of $\varv_{\rm 
turb}\simeq 8$\,km\,s$^{-1}$. The resulting fit to the $\lambda 5577$ line is 
shown as the dashed line in the bottom panel of Fig.\,\ref{wind}. 

For the other two [O{\sc i}] lines ($\lambda 6300$ and $\lambda 6364$) this
model does not hold (dotted lines in the lower panel of Fig.\,\ref{wind}).
Instead, the observed profiles are much narrower and the peak-separation is
less pronounced than in the $\lambda 5577$ line. Nevertheless, we find good 
fits to these line profiles, too, if we assume an outflow velocity of only 
16\,km\,s$^{-1}$, but a turbulent velocity of 11\,km\,s$^{-1}$ (dashed lines in 
the lower panel of Fig.\,\ref{wind}). This would imply that because these lines 
are generated much further out, the outflow velocity of the disk would have 
slightly slown down, with a simultaneous slight increase of the turbulent 
velocity.

\subsubsection{Keplerian rotating disk or ring}

\begin{figure}[t!]
\begin{center}
\resizebox{\hsize}{!}{\includegraphics{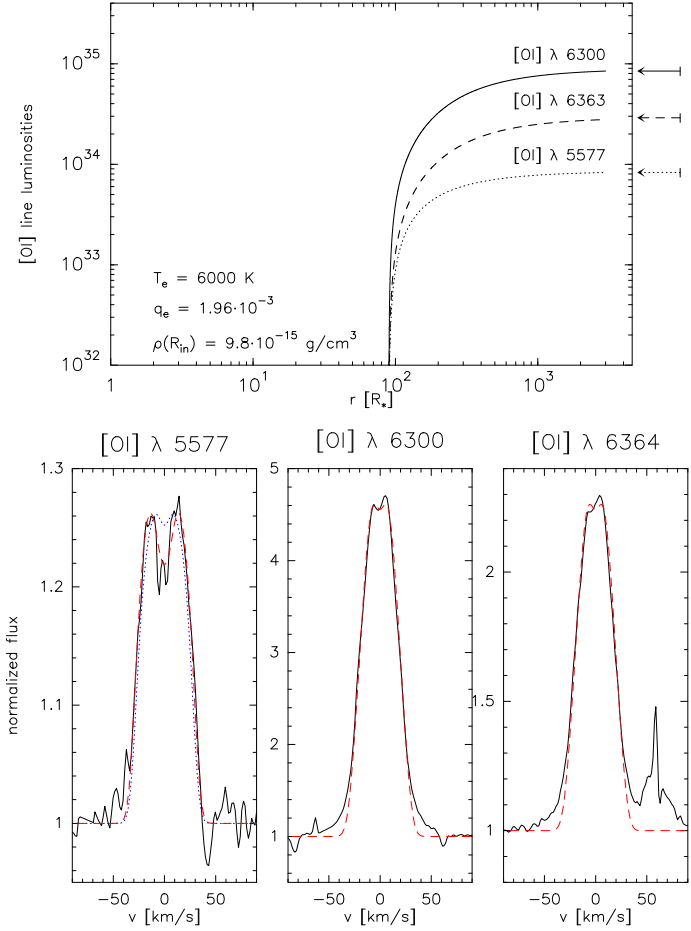}}
\caption{{\sl Top panel:} Same as Fig.\,\ref{wind} but for the Keplerian 
rotating disk scenario. {\sl Bottom panel:} Fits to the line profiles. The 
dotted profile fitted to the $\lambda 5577$ line is for a purely rotating disk 
model; the dashed lines in all profiles account for our best-fit model.}
\label{kepler}
\end{center}
\end{figure}

The decrease in peak separation seen in the [O{\sc i}] lines from a higher 
value at higher densities, i.e., closer to the star, to a lower value further 
out, could be a clear indication for Keplerian rotation. Therefore we now
calculate the [O{\sc i}] line luminosities and the profiles with a Keplerian 
rotating disk model. To translate the rotation velocities as inferred from the
peak separations of the line profiles into radial distances, we need to know
the current mass of the star. Its effective temperature and luminosity, when
compared to evolutionary tracks of rotating (Meynet \& Maeder \cite{MeMa05}) 
and non-rotating stars at SMC metallicity (Charbonnel et al. 
\cite{Charbonnel}), place S\,65 within the HR diagram to a region in agreement 
with a post-main sequence evolutionary phase of a star with an initial mass in 
the range 35--40\,$M_{\odot}$, and we assume a current mass of $\sim 
35\,M_{\odot}$. 

The peak separations of the $\lambda 6300$ and $\lambda 6364$ lines deliver a 
minimum rotation velocity of about $5\ldots 6$\,km\,s$^{-1}$, which sets the 
outer edge of the [O{\sc i}] line emission region to about $R_{\rm out}\simeq
3000$\,R$_{*}$. The $\lambda 5577$ line, which is created much closer to the 
star, thus delivers the velocity information from the inner parts of the 
Keplerian rotating disk. Due to the extended region over which the [O{\sc i}] 
line is generated, its peak separation does not reflect the 
rotation velocity at the inner edge of the disk, but the velocity at the 
distance where the bulk of emission in this line is created.  
The distance at which a rotation velocity of 22\,km\,s$^{-1}$ is achieved is at 
about 170\,R$_{*}$. The $FWHM$ of about 56\,km\,s$^{-1}$ 
implies a rotation velocity of 28\,km\,s$^{-1}$ and a distance of 105\,R$_{*}$ 
at which already some substantial fraction of the emission must be generated. 
The real inner edge of the disk must be slightly closer to the star though,
and our test calculations revealed a most likely distance of $R_{\rm in}\simeq
90$\,R$_{*}$ delivering a maximum rotation velocity to the profile of the 
$\lambda 5577$ line of $\varv_{\rm rot, max} \simeq 30$\,km\,s$^{-1}$.

While the velocity distribution in a Keplerian rotating disk is well known, the 
situation is far less clear for the radial density distribution within this 
disk. 
For simplicity we adopt a $r^{-2}$ behavior of the density distribution. A 
justification for this choice might be given by the request that the disk must 
have been formed from an outflow or ejection process obeying the mass 
continuity relation. In addition, a $r^{-2}$ density distribution is also known 
to settle over large regions within the so-called wind-wind interaction regions 
(e.g., Steffen \& Sch\"{o}nberner \cite{Steffen}). However, different power-law 
density distributions, as, e.g., found in Keplerian disks around pre-main 
sequence stars, cannot be excluded.

For a first guess of the ionization fraction and density parameter at the inner
edge of the disk, we calculate the line luminosities and the $\lambda 6300/
\lambda 5577$ line ratio for the same range of constant temperatures (5000\,K
to 7000\,K) and a disk extending from $\sim 90$\,R$_{*}$ to $\sim 
3000$\,R$_{*}$. Proceeding as in Sect.\,\ref{outflow} we find the following 
plausible ranges for the disk parameters:  
$q_{\rm e}\simeq (1.3\ldots 3.6)\times 10^{-3}$ and $F_{\rm m,
Disk}/\varv_{\rm out}\simeq (0.5\ldots 1.0)\times 10^{-10}$\,g\,cm$^{-3}$, or,
when expressed in terms of the density, $\rho$, at the inner edge of the disk,
$\rho(R_{\rm in}) \simeq (0.6\ldots 1.2)\times 10^{-14}$\,g\,cm$^{-3}$.

The results of the line luminosity calculations performed for a constant 
temperature of $T_{\rm e} = 6000$\,K are shown in the top panel of 
Fig.\,\ref{kepler}. They indicate that for a Keplerian rotating 
disk the ionization fraction, $q_{\rm e}$, must be 
higher by about a factor of 100 compared to the one found for the outflowing 
disk model. This trend is clear, because the line formation is now forced to
happen at much larger distances from the star, where the O{\sc i} density is
generally much lower. Therefore, a higher number of collision partners (i.e.,
free electrons) is necessary to produce the same line luminosity. Nevertheless, 
even with this much higher ionization fraction, the disk can still be 
considered as being neutral in hydrogen ($q_{\rm H^{+}} < 2\times 10^{-3}$).

\begin{figure}[t!]
\begin{center}
\resizebox{\hsize}{!}{\includegraphics{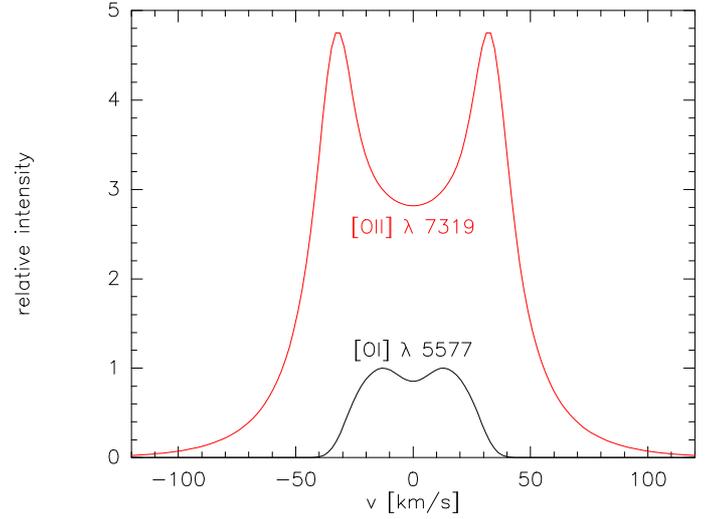}}
  \caption{Expected strength and shape of the [O{\sc ii}] $\lambda 7319$
  line in the Keplerian rotating disk scenario.}
  \label{fig5}
\end{center}
\end{figure}

The calculated profile fits to the [O{\sc i}] lines, resulting from our 
Keplerian disk model, are shown in the bottom panel of Fig.\,\ref{kepler}. 
It is obvious that the $\lambda 6300$ and $\lambda 6364$ lines can be fitted 
quite well, as shown by our dashed-line fits to these two lines. However, the 
same model fails (indicated by the dotted line) to reproduce the $\lambda 
5577$ line profile. Although the Keplerian disk model results in a slightly 
double-peaked profile for that line as well, its shape is nevertheless 
significantly different from the observed one. Especially the inter-peak 
strength is strongly overestimated. To obtain a better reproduction of the 
$\lambda 5577$ line profile, we thus have to assume that the material at the 
inner edge of the disk is not purely rotating, but an additional kinematical 
component, resulting in a double-peaked line profile, must exist as well. 
Because our investigations in the previous section have shown that an 
equatorial outflow produces double-peaked line profiles, we can speculate that 
the disk material is not revolving on stable Keplerian but rather on slightly 
outwards spiraling orbits.

Testing this hypothesis revealed that to reproduce the $\lambda 5577$
line profile we need to include an additional outflow velocity at the 
inner edge of the [O{\sc i}] line forming region of $\varv_{\rm out} = 
(9.0\pm 0.5)$\,km\,s$^{-1}$. The resulting line profile
is shown as the dashed line fit to the $\lambda 5577$ line in the lower panel
of Fig.\,\ref{kepler}. On the other hand, the profiles of the lines
$\lambda\lambda 6300, 6364$, which are formed at a much larger distance, are
not compatible with such an outflow component. In principle no outflow is 
needed for these lines, but an outflow component of $\varv_{\rm 
out} \la 2$\,km\,s$^{-1}$ would still reproduce the profiles reasonably
well, indicating a slow-down of the outflow. 

If the observed [O{\sc i}] lines result indeed from a Keplerian (or 
quasi-Keplerian) rotating disk, the question arises whether this disk extends 
down to the stellar surface, or whether the material is detached from the 
stellar surface. If the disk extends down to (or close to) the stellar surface, 
the disk parts within 90\,$R_*$, from which no contribution to the [O{\sc i}] 
emission is seen, should consequently be ionized. Because ionized oxygen is 
also known to have forbidden emission lines in the optical spectral range, we 
calculated the expected line 
luminosity and profile for the strongest of these lines, which is usually the 
[O{\sc ii}] $\lambda 7319$ line, arising in the inner disk parts that extends 
from the stellar surface out to 90\,$R_*$. The resulting line profile and 
intensity are then compared to the weakest of our [O{\sc i}] lines, i.e., the 
$\lambda 5577$ line. This comparison is shown in Fig.\,\ref{fig5}. The 
[O{\sc ii}] line turns out to be much broader (due to the higher rotation 
velocities closer to the star) and about five times more intense than the 
$\lambda 5577$ line. Therefore it should be clearly detectable. But our 
spectra do not show indications for the presence of any [O{\sc ii}] line. This 
could indeed indicate that if the 
Keplerian rotating disk scenario is the correct one, the material must be 
detached from the star.

\section{Discussion}\label{discussion}

\begin{table}[t!]
  \begin{center}
  \caption{Best-fit parameters of the two competing scenarios for S\,65.}
  \label{best_fit}
  \begin{tabular}{c|cc|cc}\hline
  \hline
  Parameter & \multicolumn{2}{c}{Outflowing disk} & \multicolumn{2}{|c}{Keplerian disk} \\  & $R_{\rm in}$ & $R_{\rm out}$ & $R_{\rm in}$ & $R_{\rm out}$ \\
  \hline
  $r [R_{*}]$ & 1 & $\ga 400$ & $\sim 90$ & $\sim 3000$ \\
  $\varv_{\rm outflow}$ [km\,s$^{-1}$] & $22\pm 1.0$ & $16\pm 1.0$ & $9\pm 0.5$ & $\la 2.0$ \\
  $\varv_{\rm rot}$ [km\,s$^{-1}$] & --- & ---  & $30\pm 1.0$ & $5\pm 0.5$ \\
  $\varv_{\rm turb}$ [km\,s$^{-1}$] & $8\pm 0.5$  & $11\pm 0.5$ & --- & ---  \\
  $n_{\rm e}(r)$ [cm$^{-3}$] & $1.2\times 10^{9}$ & $< 7.6\times 10^{3}$ & $8.2
      \times 10^{6}$ & $7.4\times 10^{3}$ \\
  \hline
\end{tabular}
\end{center}
\end{table}

\subsection{Outflowing versus Keplerian rotating disk}

So far, modeling of the [O{\sc i}] line luminosities and profiles alone did not
allow us to unambiguously distinguish between the outflowing and the Keplerian 
rotating disk model. Both can provide a plausible scenario for the neutral 
material around S\,65, and we summarized the best-fit model parameters for each
in Table\,\ref{best_fit}. The question thus arises whether we can find 
additional, predominantely kinematical hints in our data that would allow us to 
favor one of the disk models. 

Besides the [O{\sc i}] lines, our spectra display clearly double-peaked
line profiles of the [Ca{\sc ii}] lines (see Fig.\,\ref{forb}). The measured
velocities and equivalent widths are listed in Table\,\ref{CaII}.
The observed line ratio [Ca{\sc ii}] $\lambda 7291/\lambda 7324$ of $1.45\pm
0.13$ does not give a conclusive hint for the formation region of these 
lines, but agrees more with the formation in the low-density limit
(1.495) rather than in the high-density limit (1.535). These values have been
determined and discussed in detail by Hartigan et al. (\cite{Hartigan}), 
and the critical density is $\sim 5\times 10^{7}$\,cm$^{-3}$. 

The measured peak and $FWHM$ velocities are very similar to the values found
for the [O{\sc i}] $\lambda 5577$ line, indicating that the [Ca{\sc ii}] lines
originate from the same region. In both disk scenarios the electron density 
in the [O{\sc i}] line formation regions are the same, i.e., in the region of 
the [O{\sc i}] $\lambda 5577$ line formation it ranges from $n_{\rm e} = 
8.0\times 
10^{6}\ldots 1.5\times 10^{6}$\,cm$^{-3}$, while it is much lower within the 
[O{\sc i}] $\lambda\lambda 6300, 6364$ line formation region, extending 
over $n_{\rm e} = 5.0\times 10^{5}\ldots 7.5\times 10^{3}$\,cm$^{-3}$.
Consequently, the [Ca{\sc ii}] lines are indeed formed below the critical 
density as expected from the measured line ratio. Although the kinematics 
derived from the [Ca{\sc ii}] lines cannot help to distinguish between the 
outflowing and the Keplerian rotating disk we emphasize that the 
intense and clearly double-peaked [Ca{\sc ii}] lines seen in the spectrum of 
S\,65 provide an additional strong indication for the existence of a gaseous 
disk.

Nevertheless, we think that the Keplerian rotating disk is the more reasonable
scenario. Supporting kinematical arguments in favor of the Keplerian disk
are given by: (i) the symmetric, unshifted  
forbidden emission lines and (ii) the sharp absorption components.

We measured the $FWHM$ velocities of all unblended forbidden emission lines.
The velocities of all lines are very similar, resulting in a mean value of
$(37.09\pm 4.16)$\,km\,s$^{-1}$. This value is slightly lower than the $FWHM$ 
velocities of the [O{\sc i}] $\lambda\lambda 6300, 6364$ lines, indicating 
that the bulk of forbidden emission lines originates in regions near but
beyond (i.e. at larger distances and hence lower velocities than) the 
[O{\sc i}] lines. This scenario is difficult to explain with an outflowing
disk model, especially because that model requires an increase in turbulent 
velocity with distance from the star. On the other hand, it is logical to
explain the profiles of the forbidden emission lines with the Keplerian 
disk model. The symmetric line profile with no peak separation means that the 
emission is formed over a large disk, as for the [O{\sc i}] 
$\lambda\lambda 6300, 6364$ lines, but extending to distances at which the 
rotation is too small ($< 5$\,km\,s$^{-1}$) for the peaks to be still 
separated. 

\begin{table}[t!]
  \begin{center}
  \caption{Velocities obtained from the peak separation and the $FWHM$ values,
and equivalent widths of the [Ca{\sc ii}] emission lines.}
  \label{CaII}
  \begin{tabular}{cccc}\hline
  \hline
$\lambda$ & $\varv_{\rm peaks}$ & $\varv_{FWHM}$ & $EW$ \\ %& $\varv_{\rm max}$ \\
$[$\AA$]$ & [km\,s$^{-1}$] & [km\,s$^{-1}$] & $[\AA]$ \\ %& [km\,s$^{-1}$] \\
\hline
7291 & $42.0\pm 2$ & $55.2\pm 2$ & $3.04\pm 0.21$\\  % & $50.4\pm 2$\\
7324 & $38.6\pm 2$ & $56.8\pm 2$ & $2.09\pm 0.11$\\  % & $53.7\pm 2$\\
\hline
\end{tabular}
\end{center}
\end{table}

Even stronger support for the Keplerian disk scenario is given by the 
numerous shell lines, i.e., the emission lines with a sharp absorption 
component reaching below the stellar undisturbed flux. As for classical Be 
shell stars (e.g., Porter \& Rivinius \cite{Porter}), we can use these sharp 
absorption components to extract the information on the radial velocity 
structure, because these lines involve absorption of the photospheric flux 
throughout the whole disk. The wavelength displacements of the absorption 
components with respect to the line center of the emission line together with 
the measured widths of the absorption components deliver information about the 
kinematics of the absorbing medium, i.e., the kinematics of the disk.

\begin{figure}[!t]
\resizebox{\hsize}{!}{\includegraphics{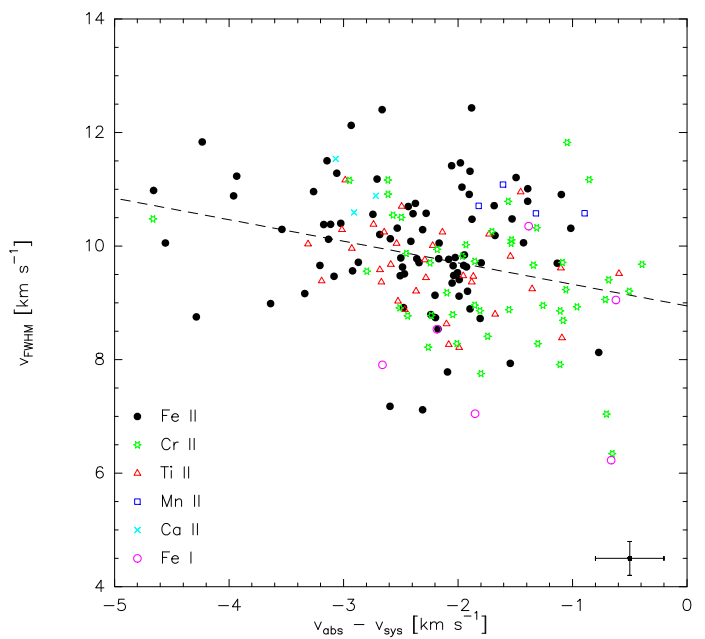}}
\caption{$FWHM$ velocities versus central peak velocities of the sharp
absorption components. Errors in velocities are indicated by the cross in the
lower right corner. The dashed line indicates the result from a linear
regression.}
\label{velocities}
\end{figure}

For all the unblended emission lines with pronounced absorption
components we measured the central wavelengths and the $FWHM$ values
of their absorption components. The central wavelengths of the absorption 
components were corrected for the systemic velocity. The velocities resulting
from their $FWHM$ values are plotted in Fig.\,\ref{velocities} versus their 
velocity offsets of their central wavelengths, defined by the difference 
$\varv_{\rm obs} - \varv_{\rm sys}$. From this plot we can draw the
following conlusions: Firstly, all absorption components are blueshifted, 
meaning that those disk parts in which the absorption happens
show a slow outflow ranging from 0 to about 5\,km\,s$^{-1}$. These low
outflow velocities agree well with those 
found for the Keplerian disk scenario (see Table\,\ref{best_fit}).
Secondly, the $FWHM$ velocities of the absorption components range from about 
6 to about 13\,km\,s$^{-1}$. Interpreting them with the 
rotation velocity of the disk, the bulk of the absorption happens at rotation 
velocities between about 3 and 6.5\,km\,s$^{-1}$. 
In addition, there seems to be a slight trend (but this trend might also
be spurious and caused by a few outliers) towards slower 
outflow for lower rotation velocities as indicated by the dashed line, which
represents the linear regression calculated for the data points. In that
respect it is also interesting to note that this trend is also seen when
comparing the results for Fe{\sc ii} with those of the (though rare) Fe{\sc i}
lines. The latter are concentrated in the right part of the plot, i.e., at the
region with lower outflow and lower rotation velocities, which is logical,
because we expect an outwards decreasing ionization structure in the disk.

A further point that needs to be discussed is the inclination of the system.
All our conclusions presented so far are based on an assumed edge-on 
orientation of the system.
As discussed in Sect.\,\ref{sys_rot} the derived projected rotation velocity is 
not conclusive and can only be considered as a lower limit to the real 
projected velocity. Even if we request the star to rotate at its critical
limit (see Sect.\,\ref{disk_form} below), we cannot claim that the inclination 
then has to follow from $i = \arcsin(\varv_{\rm rot}/\varv_{\rm crit}) = 
\arcsin(0.75) = 48.6\degr$, but it can have any value between $48.6\degr$ and 
$90\degr$. On the other hand, as mentioned in Sect.\,\ref{spec_char}, the 
numerous shell lines speak in favor of a system viewed more or less edge-on. 
With a total opening angle for B[e] supergiant stars' disks of $\sim 20\degr$ 
(see Sect.\,\ref{outflow}), a minimum inclination of about $\sim 80\degr$ might 
still be possible for the line-of-sight to pass through the disk material. In 
this case, the difference in velocity compared to a completely edge-on 
orientation is negligible. Yet we have to admit that this argument only 
holds if the disk maintains its constant opening angle, i.e. if its outer parts 
are not flared. So far nothing is known about the structure and shape (and 
especially about a possible flaring) of the disks around B[e] supergiants at 
large distances. Therefore we consider the close to edge-on orientation 
(i.e. $i\ga 80\degr$) used in our model as a plausible 
scenario.

Although the above summarized arguments can only be regarded as qualitative 
indications rather than real proofs, they nevertheless seem to speak in favor 
of a Keplerian disk scenario. What remains to be discussed is therefore the 
problem of the origin of such a disk. 

\subsection{Possible disk formation mechanism}\label{disk_form}

The source 
S\,65 is in an evolved evolutionary phase of an early-type progenitor star, 
with an initial mass of $35\ldots 40\,M_{\odot}$. Hydrodynamic calculations for 
a $35\,M_{\odot}$ star by Garc\'{i}a-Segura et al. (\cite{GarciaSegura}) have 
shown that due to its high mass-loss rate and wind speed during its 
main-sequence evolution, the stellar wind deposits sufficient energy into the 
surrounding medium to sweap up the circumstellar matter into a so-called 
interstellar (or main-sequence) bubble. When entering the supergiant phase, 
this bubble has grown in size up to a radius of about $35\ldots 40$\,pc 
(Garc\'{i}a-Segura et al. \cite{GarciaSegura}). Consequently, the disks seen 
around B[e] supergiants with sizes of only a few hundred to a few thousand AU 
cannot be pre-main sequence in origin. 

In single-star evolution, the disk-formation mechanism in evolved massive
stars is not well understood (for a review see, e.g., Owocki 
\cite{Owocki}). However, as first noted by Struve (\cite{Struve}) rapid 
(critical) stellar rotation seems to play the key role. Studies on disk 
formation around the most rapidly rotating stars known, i.e., the classical Be 
stars, revealed that their disks are most probably driven by viscosity (see, 
e.g., Porter \cite{John_visc}, Okazaki \cite{Okazaki}, Jones et al. 
\cite{Jones}), resulting in quasi-Keplerian rotating disks with a slow (few 
km\,s$^{-1}$) equatorial outflow component, in agreement with observations 
(Clark et al. \cite{Clark}). The star
S\,65 also seems to be rotating (close to) critical. Consequently, 
only a slight perturbance, perhaps in form of stellar pulsations (Owocki
\cite{Owocki}), would be sufficient to launch matter into orbit and to 
initiate a viscously driven disk. The kinematics found for the disk of S\,65 
seem to agree with a quasi-Keplerian disk with a low outflow velocity.

Other possibilities for disk formation would usually require a 
close companion, which cannot efficiently accrete the material transferred from 
the primary so that an accretion disk would form. The best known example for 
this mass transfer and disk-formation process in a massive evolved system is
certainly $\eta$\,Car (e.g. Soker \cite{Soker}). But though studied for
decades, none of the B[e] supergiants has so far been reported to have a close 
companion (e.g. Kraus et al. \cite{KBC}). In addition, neither the 
spectral energy distribution of S\,65 nor our high-quality spectra, which have 
been taken on different dates covering observing periods as long as 11 years 
and as short as 24 hours, show any indication for a companion. Also, our 
spectra show no evidence for ongoing accretion. Although the existence of a 
companion cannot definitely be ruled out, we consider it as the less probable 
scenario for the disk formation around S\,65.

Whether the disk around S\,65 is indeed driven viscously, as suggested by the
observed (quasi-)Keplerian rotation and the slow outflow component, needs to be 
studied in a future work. Also, whether the disk indeed slows down from inside 
out as suggested by our data, is an interesting question that needs to be
investigated in much more detail than can be done here.

\subsection{The connection with LBVs}

Another still unsolved problem is related to the question whether B[e] 
supergiants are a special evolutionary phase in single-star evolution
through which only massive stars 
with very specific initial conditions will evolve, or whether all massive stars 
have to go through the B[e] supergiant phase. And if the latter is true, then 
what are the direct progenitors/descendants of the B[e] supergiant phase ?

B[e] supergiants are often discussed in connection with Luminous Blue Variables 
(LBVs). This is certainly because LBVs also show an 
indication for a disk or ring structure of circumstellar material. 
Although a bipolar or multipolar structure seen in several LBVs
like in $\eta$\,Car can be caused by binary interaction (Soker 
\cite{Soker}), the majority of LBVs have not been reported to have a close
companion. Instead, their often seen ring-shaped nebulae (e.g., Nota et al. 
\cite{Nota}) can easily be explained by wind-wind interaction in single-star 
evolution (Garc\'{i}a-Segura et al. \cite{GarciaSegura2}). 
Morphological studies of LBV nebulae imply that the slow and dense wind arises 
mainly from the equatorial regions (Nota et al. \cite{Nota}). And as for
the B[e] supergiants, rapid stellar rotation is thought to play a
major role in triggering this non-spherical mass-loss.

The second common characteristic is related to the outflow velocity.
Weis (\cite{Weis03}) studied the kinematics of LBV nebulae 
and found that their expansion velocities seem to decrease with metallicity.
The LBV nebulae in the SMC are thus expected to have expansion velocities of
less than 10\,km\,s$^{-1}$; hence S\,65 with its outflow component of a few
km\,s$^{-1}$ would fit in perfectly. With regard to this, we
add that for the B[e] supergiant R\,126 in the LMC a disk
outflow velocity of $\sim 11.5$\,km\,s$^{-1}$ was determined 
(Kraus et al. \cite{KBA}), in good agreement with the outflow velocities 
of LBV nebulae in the LMC. We also find that the gas masses in the 
disks around S\,65 ($\ga 1.5\times 10^{-2}\,M_{\odot}$) and R\,126 ($\ga 
6\times 10^{-2}\,M_{\odot}$) fit nicely to the range of gas masses found for 
LBV nebulae (Nota et al. \cite{Nota}; Weis \cite{Weis03}).

\begin{figure}[!t]
\resizebox{\hsize}{!}{\includegraphics{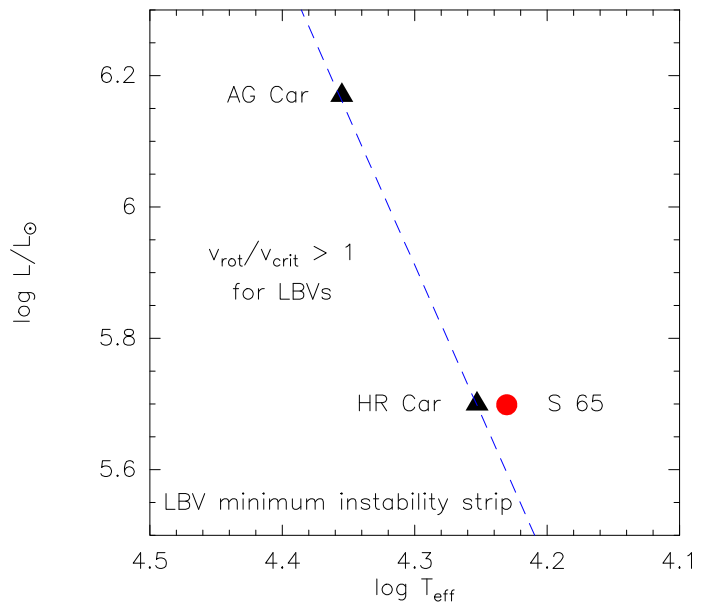}}
\caption{Location of S\,65 within the HR diagram and with respect to two 
rapidly rotating Galactic LBVs close to the LBV minimum instability strip 
(after Groh et al. \cite{Groh}). For details see text.}
\label{lbv}
\end{figure}

The last common characteristic refers to the stellar rotation speed. Groh et 
al. (\cite{Groh}) recently found for two of the Galactic LBVs, AG\,Car and 
HR\,Car, rotation velocities during their visual minimum phase in excess of 
85\% of their critical velocity. Hence Groh et al. (\cite{Groh}) defined an LBV 
minimum-instability strip, which indicates the position in the HR diagram at 
which critical rotation for LBVs is expected. This is shown in Fig.\,\ref{lbv}.
To the left of the instability strip is the physically unstable region for 
LBVs, labeled with $\varv_{\rm rot}/\varv_{\rm crit} > 1$. We also included the 
position of S\,65 into this plot. Its location is in the valid range for LBVs, 
but also very close to the instability strip.  

In summary, the rapid rotation, the close vicinity in the HR diagram to the 
LBVs and the LBV minimum instability strip, as well as the similar disk mass 
and outflow velocity derived for S\,65 might all hint towards a connection 
between LBVs and S\,65 (or perhaps B[e] supergiants in general). The most 
logical conclusion is thus that S\,65 is currently in the LBV visible 
minimum phase, after experiencing one (or more) eruptions, in which a large 
amount of material was ejected from the rapidly rotating star, predominantly in 
equatorial direction, which formed the Keplerian ring or disk-like structure. 
If this scenario is correct, then S\,65 could be in a state just 
before the new, fast wind sets in, which will compress the disk/ring material 
into the typical LBV nebula. The signatures of Keplerian rotation, as still 
seen in the disk of S\,65, might disappear during this wind compression 
process, because the material will be radially accelerated outwards.

We cannot conclusively answer whether S\,65 indeed belongs to the class of 
LBVs, especially because a star is called an LBV only after a giant eruption 
and the typical S\,Dor variability has been recorded. Nevertheless, there is 
ample evidence that S\,65 shares several of the major LBV characteristics.

%______________________________________________________________

\section{Conclusions}

We studied the circumstellar material of the rapidly rotating SMC B[e] 
supergiant star S\,65. The sharp absorption components seen in its numerous 
emission lines indicate a high-density disk viewed almost edge-on. 
Surprisingly, the spectrum also displays two very intense
and clearly double-peaked [Ca{\sc ii}] lines, supporting the high-density
gaseous disk hypothesis. Based on the modeling of both, the observed line 
luminosities of the [O{\sc i}] lines and their line profiles, we found that the 
disk around the B[e] supergiant S\,65 must be predominantly neutral in 
hydrogen. The total amount of ionized hydrogen is found to be less than 
$\sim 0.1$\%, and the total amount of gas traced within the [O{\sc i}] 
line-forming region is $\sim 1.5\times 10^{-2}\,M_{\odot}$. In combination with 
the kinematical information provided by the numerous lines, our model results 
favor the interpretation of a detached (quasi-)Keplerian rotating disk rather 
than an 
outflowing disk. The location of S\,65 in the HR diagram places the star
close to the recently defined LBV minimum instability strip. Together with its 
high rotation velocity close to the critical limit and its large amount of 
slowly expanding circumstellar material similar to that seen for LBV nebulae,
it seems that S\,65 might be the link between B[e] supergiants and LBVs.

%______________________________________________________________

\begin{acknowledgements}
We thank the referee, Dr. J.H. Kastner, for his valuable 
comments on the manuscript.
This research made use of the NASA Astrophysics Data System (ADS).
M.K. acknowledges financial support from GA\,AV \v{C}R number KJB300030701. 
M.B.F. acknowledges financial support from the Centre National de la Recherche 
Scientifique (CNRS) and from the Programme National de Physique Stellaire.
M.B.F. also acknowledges Conselho Nacional de Desenvolvimento Cient\'{i}fico
e Tecnol\'{o}gico (CNPq-Brazil) for the post-doctoral grant.
\end{acknowledgements}


\begin{thebibliography}{}


\bibitem[1993]{Charbonnel}
        Charbonnel C., Meynet G., Maeder A., Schaller, G., \& Schaerer, D. 
        1993, A\&AS, 101, 415
\bibitem[2003]{Clark}
        Clark, J. S., Tarasov, A. E., \& Panko, E. A. 2003, A\&A, 403, 239
\bibitem[2004]{Cure04}
        Cur\'{e}, M. 2004, ApJ, 614, 929
\bibitem[2005]{Cure05}
        Cur\'{e}, M., Rial, D. F., \& Cidale, L. 2005, A\&A, 437, 929
\bibitem[2005]{Dufton}
        Dufton, P. L., Ryans, R. S. I., Trundle, C., et al. 2005, A\&A, 434,
        1125
\bibitem[1960]{Feast}
        Feast, M. W., Thackery, A. D., \& Wesselink, A. J. 1960, MNRAS, 121, 
        344
\bibitem[1996a]{GarciaSegura2} 
        Garc\'{i}a-Segura, G., Mac Low, M.-M., \& Langer, N. 1996a, A\&A, 305,
        229
\bibitem[1996b]{GarciaSegura} 
        Garc\'{i}a-Segura, G., Langer, N., \& Mac Low, M.-M. 1996b, A\&A, 316,
        133
\bibitem[1976]{Gray}
        Gray, D. F. 1976, The observation and analysis of stellar photospheres
        (New York: Wiley)
\bibitem[1998]{Grevesse} 
        Grevesse, N. \& Sauval, A. J., 1998, Space Sci. Rev. 85, 161  
\bibitem[2009]{Groh}
        Groh, J.H., Damineli, A., Hillier, D.J., et al. 2009, ApJ, 705, L\,25 
\bibitem[2000]{Hanuschik}
        Hanuschik, R. W. 2000, in: The Be Phenomenon in Early-Type Stars, ed.
        M. A. Smith, H. F. Henrichs, \& J. Fabregat (San Francisco: ASP),
        ASP Conf. Ser. Vol. 214, 518
\bibitem[1994]{Hamuy} 
        Hamuy, M., Suntzeff, N. B., Heathcote, S. R., et al. 
        1994, PASP, 106, 566
\bibitem[2004]{Hartigan}
        Hartigan, P., Edwards, S., \& Pierson, R., 2004, ApJ, 609, 261 
\bibitem[2008]{Jones}
        Jones, C. E., Sigut, T. A. A., \& Porter, J. M. 2008, MNRAS, 386, 1922
\bibitem[1980]{Kafatos}
        Kafatos, M., \& Lynch, J. P. 1980, ApJS, 42, 611
\bibitem[2006]{Kastner}
        Kastner, J. H., Buchanan, C. L., Sargent, B., \& Forrest, W. J. 2006,
        ApJ, 638, L\,29
\bibitem[2006]{Keller}
        Keller, S. C., \& Wood, P. R. 2006, ApJ, 642, 834
\bibitem[2006]{Kraus06} Kraus, M. 2006, A\&A, 456, 151
\bibitem[2005]{KB} Kraus, M., \& Borges Fernandes, M. 2005, in:
        The Nature and Evolution of Disks Around Hot Stars, ed. R. Ignace \&
        K.G. Gayley (San Francisco: ASP), ASP Conf. Ser. Vol. 337, 254
\bibitem[2003]{KL}
        Kraus, M., \& Lamers, H. J. G. L. M., 2003, A\&A, 405, 165
\bibitem[2007]{KBA} 
        Kraus, M., Borges Fernandes, M., \& de Ara\'ujo, 
        F. X. 2007, A\&A, 463, 627
\bibitem[2010]{KBC}
        Kraus, M., Borges Fernandes, M., \& Chesneau, O. 2010, in 
        Binaries, Key to Comprehension of the Universe, ed. A. Pr\v{s}a \&
        M. Zejda, ASP Conf. Ser., in press 
\bibitem[2008]{Kraus08} 
        Kraus, M., Borges Fernandes, M., Kub\'at, J., \& de 
        Ara\'ujo, F. X. 2008, A\&A, 487, 697
\bibitem[2006]{Vlieland}
        Kraus, M., Borges Fernandes, M., Andrade Pilling, D., \& de Ara\'ujo, 
        F.X., 2006, in Stars with the B[e] Phenomenon, ed. M. Kraus \& A.S.
        Miroshnichenko (San Francisco: ASP), ASP Conf. Ser. Vol. 355, 163 
\bibitem[1991]{LP}
        Lamers, H. J. G. L. M., \& Pauldrach, A. W. A. 1991, A\&A 244, L5
\bibitem[2000]{MaMe00}
        Maeder, A., \& Meynet, G. 2000, A\&A, 361, 159
\bibitem[1992]{Magalhaes}
        Magalh\~aes, A. M. 1992, ApJ, 398, 286
\bibitem[2006]{Magalhaesetal}
        Magalh\~aes, A. M., Melgarejo, R., Pereyra, A., \& Carciofi, A. C.
        2006, in Stars with the B[e] Phenomenon, ed. M. Kraus \& A. S. 
        Miroshnichenko (San Francisco: ASP), ASP Conf, Ser., 355, 147
\bibitem[1976]{Maurice}
        Maurice, E. 1976, A\&A, 50, 463
\bibitem[1988a]{McGregor_a}
        McGregor, P. J., Hillier, D. J., \& Hyland, A. R. 1988a, ApJ, 334, 639
\bibitem[1988b]{McGregor_b}
        McGregor, P. J., Hyland, A. R., \& Hillier, D. J. 1988b, ApJ, 324, 1071
\bibitem[1989]{McGregor89}
        McGregor, P. J., Hyland, A. R., \& McGinn, M. T. 1989, A\&A, 223, 237
\bibitem[2001]{Melgarejo}
        Melgarejo, R., Magalh\~aes, A. M., Carciofi, A. C. \& Rodrigues, C. V.
        2001, A\&A, 377, 581
\bibitem[1983]{Mendoza} 
        Mendoza, C. 1983, IAU Symp., 103, 143
\bibitem[2005]{MeMa05} 
        Meynet, G., \& Maeder, A. 2005, A\&A, 429, 581
\bibitem[1996]{Morris}
        Morris, P. W., Eenens, P. R. J., Hanson, M. M., Conti, P. S., \& 
        Blum, R. D. 1996, ApJ, 470, 597
\bibitem[1995]{Nota}
        Nota, A., Livio, M., Clampin, M., \& Schulte-Ladbeck, R., 1995, ApJ,
        448, 788
\bibitem[1981]{Muratorio}
        Muratorio, G. 1981, A\&AS, 43, 111
\bibitem[2001]{Okazaki}
        Okazaki, A. T. 2001, PASJ, 53, 119 
\bibitem[2006]{Owocki} 
        Owocki, S.P. 2006, in Stars with the B[e] Phenomenon, ed. M. Kraus \&
        A. S. Miroshnichenko (San Francisco: ASP), ASP Conf, Ser., 355, 219 
\bibitem[2000]{Pelupessy}
        Pelupessy, I., Lamers, H. J. G. L. M., \& Vink, J.S. 2000, 
        A\&A 359, 695
\bibitem[1999]{John_visc}
        Porter, J. M. 1999, A\&A, 348, 512
\bibitem[2003]{John}
        Porter, J. M. 2003, A\&A, 398, 631
\bibitem[2003]{Porter}
        Porter, J. M., \& Rivinius, Th. 2003, PASP, 115, 1153
\bibitem[1984]{Prevot}
        Pr\'evot, M. L., Lequeux, J., Maurice, E., et al. 1984, A\&A, 132, 389
\bibitem[2003]{Soker}
        Soker, N. 2003, ApJ, 597, 513
\bibitem[2001]{Stahl}
        Stahl, O. 2001, in Eta Carinae \& Other Mysterious Stars
        ed. T. Gull, S. Johansson \& K. Davidson
        (San Francisco: ASP), ASP Conf. Ser., 242, 163
\bibitem[2003]{Steffen}
        Steffen, M., \& Sch\"{o}nberner, D. 2003, in Planetary Nebulae: Their
        Evolution and Role in the Universe, ed. S. Kwok, M. Dopita, \& R. 
        Sutherland (San Francisco: ASP), IAU Symp. Vol. 209, 439
\bibitem[2000]{Stoerzer}
        St\"{o}rzer, H., \& Hollenbach, D. 2000, ApJ, 539, 751
\bibitem[1931]{Struve}
        Struve, O. 1931, ApJ, 73, 94
\bibitem[2004]{Townsend}
        Townsend, R.H.D., Owocki, S.P., \& Howarth, I.D., 2004, MNRAS, 350, 189
\bibitem[2003]{Weis03}
        Weis, K., 2003, A\&A, 408, 205 
\bibitem[1966]{Wiese} 
        Wiese, W. L., Smith, M. W., \& Glennon, B. M. 1966, Atomic Transition 
        Probabilities, Vol. 1 (National Standard Reference Data System, 
        Washington D.C.)
\bibitem[1989]{Zickgraf89}
        Zickgraf, F.-J. 1989, in Physics of Luminous Blue Variables, ed.
        K. Davidson, A. F. J. Moffat, \& H. J. G. L. M. Lamers (Dordrecht:
        Kluwer Academic Publishers), IAU Colloquium 113, 117 
\bibitem[2000]{Zickgraf00}
        Zickgraf, F.-J. 2000, in The Be Phenomenon in
        early-type stars, ed. M. A. Smith, H. F. Henrichs, \& J. Fabregat
        (San Francisco: ASP), ASP Conf. Ser., 214, 26
\bibitem[1985]{Zickgraf85}
        Zickgraf, F.-J., Wolf, B., Stahl, O., Leitherer, C., \& Klare, G.
        1985, A\&A, 143, 421
\bibitem[1986]{Zickgraf86}
        Zickgraf, F.-J., Wolf, B., Stahl, O., Leitherer, C., \& Appenzeller, I.
        1986, A\&A, 163, 119
\bibitem[1989]{Zick89}
        Zickgraf, F.-J., Wolf, B., Stahl, O., \& Humphreys, R. M.
        1989, A\&A, 220, 206

\end{thebibliography}
\end{document}